\documentclass{tCPH2e}

\usepackage{epstopdf}
\usepackage{subfigure}
\usepackage{amssymb}

\theoremstyle{plain}

\usepackage{color}
\definecolor{mymagenta}{cmyk}{0,1,0,0.12}

\theoremstyle{definition}

\theoremstyle{remark}

\begin{document}


\articletype{Review}

\title{Lattice gauge theories simulations in the quantum information era}

\author{
\name{M. Dalmonte\textsuperscript{a}
and S. Montangero\textsuperscript{b}$^{\ast}$\thanks{$^\ast$Corresponding author. Email: simone.montangero@uni-ulm.de}}
\affil{\textsuperscript{a} Institute for Theoretical Physics, University of Innsbruck \&
Institute for Quantum Optics and Quantum Information of the Austrian Academy of Sciences, A-6020 Innsbruck, Austria;\\
\textsuperscript{b}Institute for complex quantum systems \& Center for Integrated Quantum Science and Technologies, Ulm university, Albert-Einstein-Allee 11, 89075 Ulm, Germany.}
\received{v4.0 released January 2015}
}

\maketitle

\begin{abstract}
{The many-body problem is ubiquitous in the theoretical description of physical phenomena, ranging from the behavior of elementary particles to the physics of electrons in solids.} Most of our understanding of many-body systems comes from analyzing the symmetry properties of Hamiltonian and states: the most striking example are gauge theories such as quantum electrodynamics, where a local symmetry strongly constrains the microscopic dynamics. The physics of such gauge theories is relevant for the understanding of a diverse set of systems, including frustrated quantum magnets and the collective dynamics of elementary particles within the standard model. In the last few years, several approaches have been put forward to tackle the complex dynamics of gauge theories using quantum information concepts. In particular, quantum simulation platforms have been put forward for the realization of synthetic gauge theories, and novel classical simulation algorithms based on quantum information concepts have been formulated. In this review we present an introduction to these approaches, illustrating the basics concepts and highlighting  
the connections between apparently very different fields, and report the recent developments in this new thriving field of research.
\end{abstract}


\section{Introduction}

Many-body problems, and in particular quantum ones, lie at the border of our current knowledge in many different field of science, 
from very applied ones, as the electronic structure problem with impact on chemistry and drugs research~\cite{Szabo1996},  
the development of quantum technologies~\cite{Nielsen2011,Kurizki2015}, or the quest for understanding 
and engineering new material properties, e.g. new light harvesting materials that might impact our everyday life~\cite{Cheng2009,Scholes2011};  
to basic research such as high energy physics which aims to understand the fundamental constituents 
of the universe (the Standard Model)~\cite{Buchmuller2006} or condensed matter physics where the comprehension of the processes underlying 
high-temperature superconductivity is still incomplete~\cite{Lee2006,Mann2011}. 

The difficulty of the many-body problem stems from the fact that combining different objects, each of them characterized 
by $d$ possible different states $\{\alpha\}$, gives raise to an exponential growth of the number of possible configurations for the whole 
system. Indeed, $N$ independent bodies with $d$ states each counts for $d^N$ possible states 
for the whole many-body system. This is due to the fact that the components of the system are {\it a priori} independent, and thus
for any possible state of one component the others might be in any other possible combinations, that is, the 
dimension of the space of the possible configurations is the product of the individual dimensions. 
For example, the possible configurations of two six-face dices are thirtysix, while for ten dices we have already more than 
sixty millions configurations. Notice that this would not be true for correlated states: in our example, if the two dices are 
biased to give always the same number (perfect classical correlation) the dimension of the space of the possible configurations is 
only six.

In quantum mechanics, the fundamental object of interest is the wave function of the system, the amplitude of probability distribution 
$\psi(\alpha_{1} \dots \alpha_{N})$, whose modulus square gives the probability of the system to be in a $N$-body configuration 
$\alpha_{1} \dots \alpha_{N}$: that is, the object of interest is the vector containing the amplitude of probability 
for each possible system configuration. Typically, one has to update the vector of probabilities to follow the system time evolution 
or to find the eigenvector corresponding to the ground state of the system. 
These tasks quickly become impossible even on the largest available supercomputers: Indeed, having in mind the goal of 
describing a material starting from its atomic constituents and assuming for the sake of simplicity that each atom has 
only two possible relevant configurations (for example a spin up or down, or an electron in two possible orbitals),  the dimension of the configuration space 
of a cube with four atoms per edge is $2^{64}$, equal to about $10^{19}$ possible configurations. 
Given that the sum of the performance of the first five hundred nonclassified supercomputers on earth is currently 
of the order of hundreds of PFlops/sec~\cite{top500}, simply updating such vector would require some tens of seconds using all of them, or 
in alternative, some minutes on one of the most powerful supercomputer. Moreover, this example consider only a minimal structure of $64$ atoms, 
far away from what would be necessary to describe any macroscopic sample of material composed of order Avogadro number atoms!

Given the above scenario, describing the state of a many-body quantum system appears to be out of reach if no analytical solutions exist. However, a number of approximate analytical and numerical methods have been developed to describe faithfully 
the usually small  portion of interest of the exponentially big Hilbert space: they range from renormalization group (RG) methods (which rely on 
a truncation in a particular energy sector of the system)~\cite{Wilson1975}, to mean field and beyond mean field methods (which disregard or take into accounts 
only a small part of the quantum correlations of the system)~\cite{Hatting2009}, to those that computes the property of interest by means of sophisticated 
statistical samplings, mostly going under the name of Monte Carlo methods~\cite{rubinstein2011}. Of particular interest for what follows is a class of 
numerical methods which stemmed from RG methods in the context of condensed matter systems~\cite{white1992} and then has been extensively 
developed within the quantum information community: tensor network (TN) methods. As explained in details below, TN methods provide a recipe to describe 
efficiently many-body quantum systems with a controlled level of approximation accommodating as much as possible information as a 
function of the used resources~\cite{Schollwock2011}. An interesting feature of TN methods is that they complements Monte Carlo methods as they do not suffer 
from their major limitation, namely the sign problem. Thus they allow investigating scenarios -- typically out of equilibrium phenomena or at finite chemical potential -- 
where the convergence of Monte Carlo methods is severely limited~\cite{Troyer2005}. 

\begin{figure}[t]
\begin{center}
\resizebox*{8cm}{!}{\includegraphics{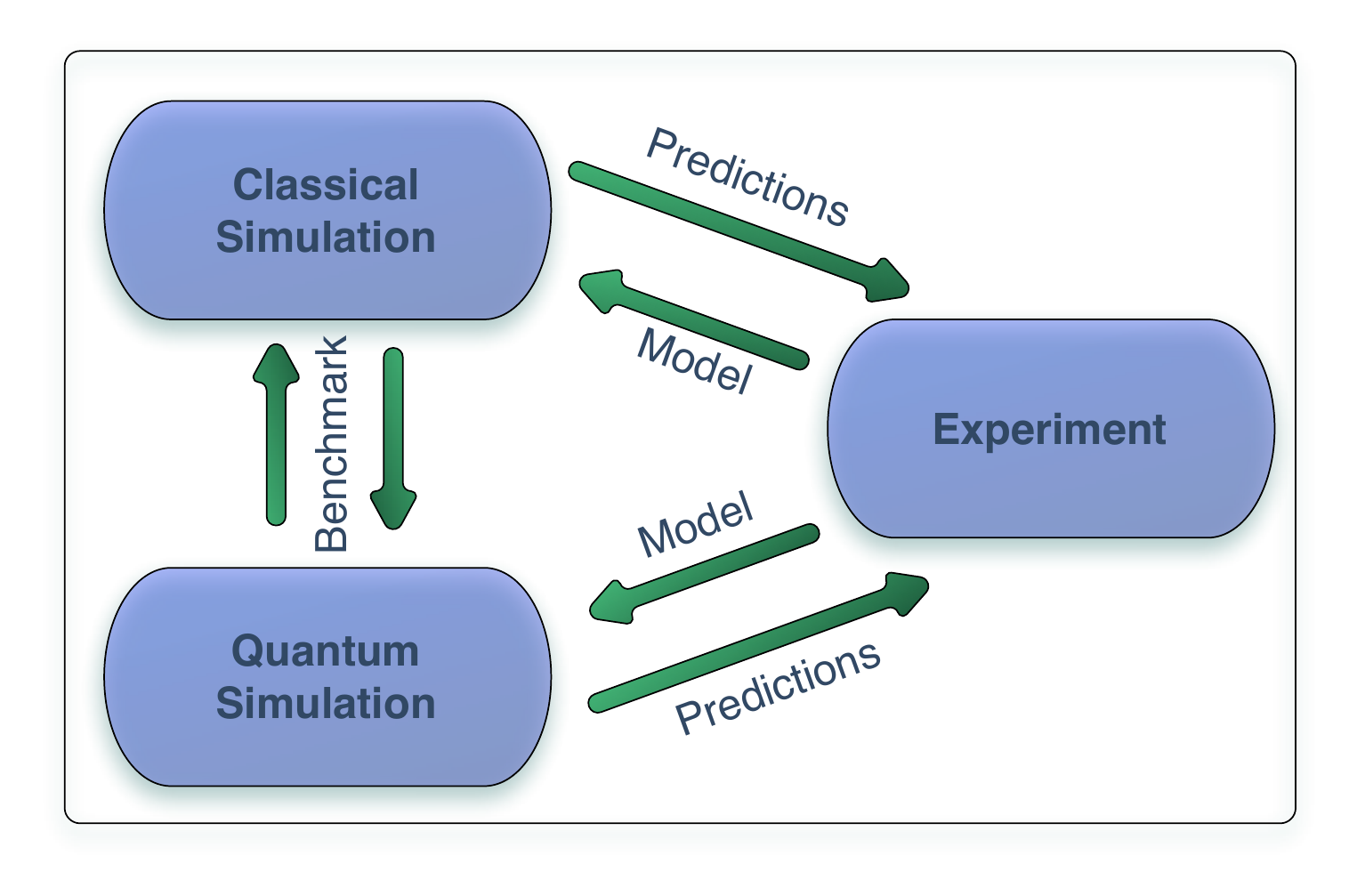}}
\caption{Three independent approaches to many-body quantum systems, and in particular to Lattice Gauge Theories, investigations: experiments in the lab, and 
classical or quantum simulations will provide information on the system of interest. The three approaches will validate theoretical models and predictions, and 
benchmark and build trust on each others.}
\end{center}
\label{investigations}
\end{figure}

An independent approach to solve the many-body problem has been put forward by R.~Feynman in the seminal work 
where he introduced the concept of a quantum simulation~\cite{Feynman1982}: he pointed out that the most natural way to simulate and study 
an unknown quantum many-body process is to use another many-body quantum system that can be controlled to such a level as to 
mimic the physics of the system under study. Indeed, the exponential growth of the Hilbert space of the system under study is automatically 
matched by another many-body quantum system. In modern words, his proposal was to develop a dedicated quantum computer or emulator able to 
simulate an interesting physical systems. Since then, quantum information science and our experimental capabilities 
have been developed to the level that different platforms (cold atoms, trapped ions, superconducting qubits, etc.)~\cite{Cirac2012} are now able to perform quantum simulations 
of several interesting many-body phenomena~\cite{Bloch2012,blatt2012quantum}. However, despite having obtained impressive achievements and performed fascinating experiments, 
quantum simulators cannot yet be used routinely to investigate unchartered territory as the so called quantum supremacy has not been proven yet~\cite{Preskill2012}. 
Nevertheless, we can envisage that investigations on many-body quantum systems will be performed in three independent ways in the future, as depicted in Fig.~\ref{investigations}: either performing an experiment to directly interrogate the system of interest, or performing a  
simulation on a classical computer or finally performing a quantum simulation in a controlled environment. 

Given that each of these approaches has pros and cons -- in terms of complexity and resources needed for their realization, reproducibility, 
access to different direct or indirect information -- most probably the future research will be based on all of them, each exploiting 
its own peculiarities and complementing and certifying the results of the others. Indeed, as it already happens nowadays, 
each approach requires a level of sophistication calling for an independent check of the results. 

As we already mentioned, small scale quantum simulations have been demonstrated for a large class of models and systems, in particular 
for a very important class playing a major role in condensed matter physics, such as interacting spins, bosons or fermions on a lattice
that model magnetic, conducting properties of materials and different phases of matter~\cite{Bloch2008,Bloch2012}. These systems can be described in general 
via Hamiltonians which can be written as
\begin{equation}
H = \sum_{i=1}^N H_i + \sum_{<i,j>} H_{i,j},
\label{ham}
\end{equation} 
where the $H_i $ represent the local Hamiltonian of each particle, spin or lattice site, the $H_{i,j}$ describe the interaction among them, and the sum  $<i,j>$
runs over all pairwise interactions present in the system (which might depend on the dimensionality of the system or on the coordination number of the lattice). 
Notice, that the Hamiltonian~\eqref{ham} is defined over discrete indices, that is, we have implicitly introduced a space discretization, that is a point lattice. 
Paradigmatic examples of such classes of Hamiltonians are short-range interacting models such as the Ising and Heisenberg models for spin systems and the 
Hubbard or Bose-Hubbard model describing particles on a lattice, as electrons hopping in an underlaying crystalline structure~\cite{Jaksch2005a,Lieb1993a}. 
However, also systems with long range interactions have been engineered, exploiting Coulomb interaction of trapped ions and Rydberg atom simulators 
are nowadays subject to intense research activity~\cite{Buchler2005,Weimer2010,Tagliacozzo2013,Glaetzle2014,Schauss2015a}.

Despite the enormous progresses experienced in the last decade, there is still an important class of phenomena that are still beyond reach of 
the quantum simulator capabilities: lattice gauge theories (LGTs) have not been implemented yet, even though in the last years an intense theoretical 
activity has started to explore the possibility of encompassing this class in the portfolio of phenomena that can be studied by means of quantum  
technologies~\cite{Wiese2014,Zohar2015a}.

\begin{figure}
\begin{center}
\begin{minipage}{160mm}
\resizebox*{16cm}{!}{\includegraphics{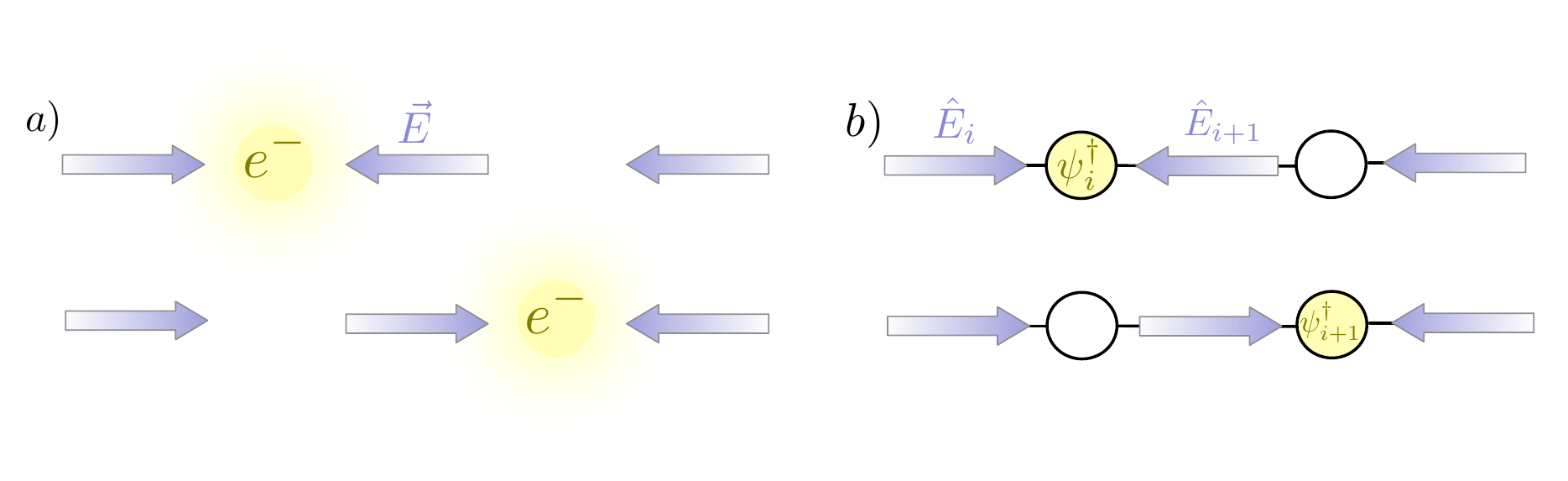}}
\hspace{6pt}
\caption{a) Gauss' law in electromagnetism: an electron $e^-$ generates an electric field $\vec E$ pointing inwards (upper configuration), 
if the electron moves in another position, the electric field changes accordingly (lower configuration). b) Lattice QED equivalent description: 
the electron lives on the vertex and the electric field on the link, the difference between the electric fields across a link obeys to Eq.~\eqref{qgauss}: they are 
equal if there is no electron on the vertex, differ by one quanta of excitation if an electron is present. 
} \label{QED}
\end{minipage}
\end{center}
\end{figure}

Gauge theories are defined promoting the global symmetry of the system to a local one, that is, from a global conservation of charge to a local one, to be satisfied 
at each point in space (or at each lattice site if the theory is defined on a lattice). A most striking example is electromagnetism. Indeed, Gauss' Law which 
relates the divergence of the electric field $\vec E$ and the charge $\rho$ at every point in space $\vec x$:
\begin{equation}
\nabla \cdot \vec E (\vec x)= \rho(\vec x),
\label{dGL}
\end{equation}
is nothing more than a constraint -- a symmetry -- to be fulfilled at every point is space. Similarly, in the quantum theory of electromagnetism, 
quantum electrodynamics (QED), we expect that such a constraint remains valid. Hereafter, for simplicity we specialize
to one-dimensional systems on a lattice: it is straightforward to generalize to arbitrary dimensions. In QED, one has two different 
degrees of freedom: the matter field $\psi$ describing the electrons, of fermionic nature, and the electric field $\hat E$ -- 
a bosonic quantum field representing the quantized excitations of the electric field, i.e. the photons. 
Differently from the previous examples, where the particles live at each lattice site, we shall specify where the two different degrees 
of freedom - the bosonic and the fermionic one - live. Indeed, we can safely define the matter field to live at the lattice site, while the electric 
field (and in general, any gauge field) lives on the link connecting two lattice sites (see Fig.~\ref{QED}).
Finally, if a Gauss' law should hold also in QED, the following condition shall be imposed:
\begin{equation}
\hat E_i - \hat E_{i+1} = \psi^\dagger_i \psi_i,
\label{qgauss}
\end{equation}
where $\hat E_i - \hat E_{i+1}$  is the change of electric field across a vertex of the lattice where charges might be present and quantified by the 
operator $\hat n_i = \psi^\dagger_i \psi_i$, which counts the number of electrons at site $i$. Finally, as we will review in detail later on, one can show that 
this gauge symmetry is again a $U(1)$ symmetry as the Hamiltonian is invariant under a phase change. However, differently from before, the phase 
change can be different in every lattice site not a global one, i.e. $q$ is an explicit function of the lattice site index $i$ for theories on the lattice and of the position in space $x$ for theories defined on the continuum. 

LGTs have been traditionally studied via Monte Carlo or directly via experiments. However, as already mentioned, in the last years 
the application of quantum information concepts to these problems has started. 
The additional theoretical and experimental 
difficulties to overcome with respect to the standard quantum simulations are that two different degrees of freedom have to 
be encoded and that local constraints of the kind given in Eq.~\eqref{qgauss} have to be implemented. Currently, 
proposals for quantum simulations of LTG in different platforms such as cold atoms in optical lattice, 
trapped ions, Rydberg atoms, and superconducting qubits have been put forward and some proof-of-principle experimental realizations have been planned~\cite{Zohar2015a,Tagliacozzo2013,Glaetzle2014,Glaetzle2015,Egusquiza2015,Barrett2013,Wiese2014,Hauke2013,Banerjee2013}. 
Alongside, tensor network methods have been applied to studies which can complement Monte Carlo ones. Indeed, it has already been 
shown that it is possible to study efficiently via TN methods Abelian and non Abelian LGTs, the real time evolution of string breaking phenomena, systems at finite 
chemical potential, in some cases obtaining results comparable with state of the art results, e.g., for the mass gaps of the massive Schwinger model~\cite{Rico2014,Pichler2015,Tagliacozzo2014,Banuls,Banuls2015,Buyens2014}. 
Even though the numerical tools and the presented results are not yet mature and mostly confined to low-dimension models, their potential encourages 
further activities to unveil their whole potential, motivated also by the fact that, despite that the problems are very challenging numerically, there are no fundamental limitations  -- such as 
the sign problem for Monte Carlo approaches -- in sight. 

\begin{figure}[t]
\begin{center}
\resizebox*{8cm}{!}{\includegraphics{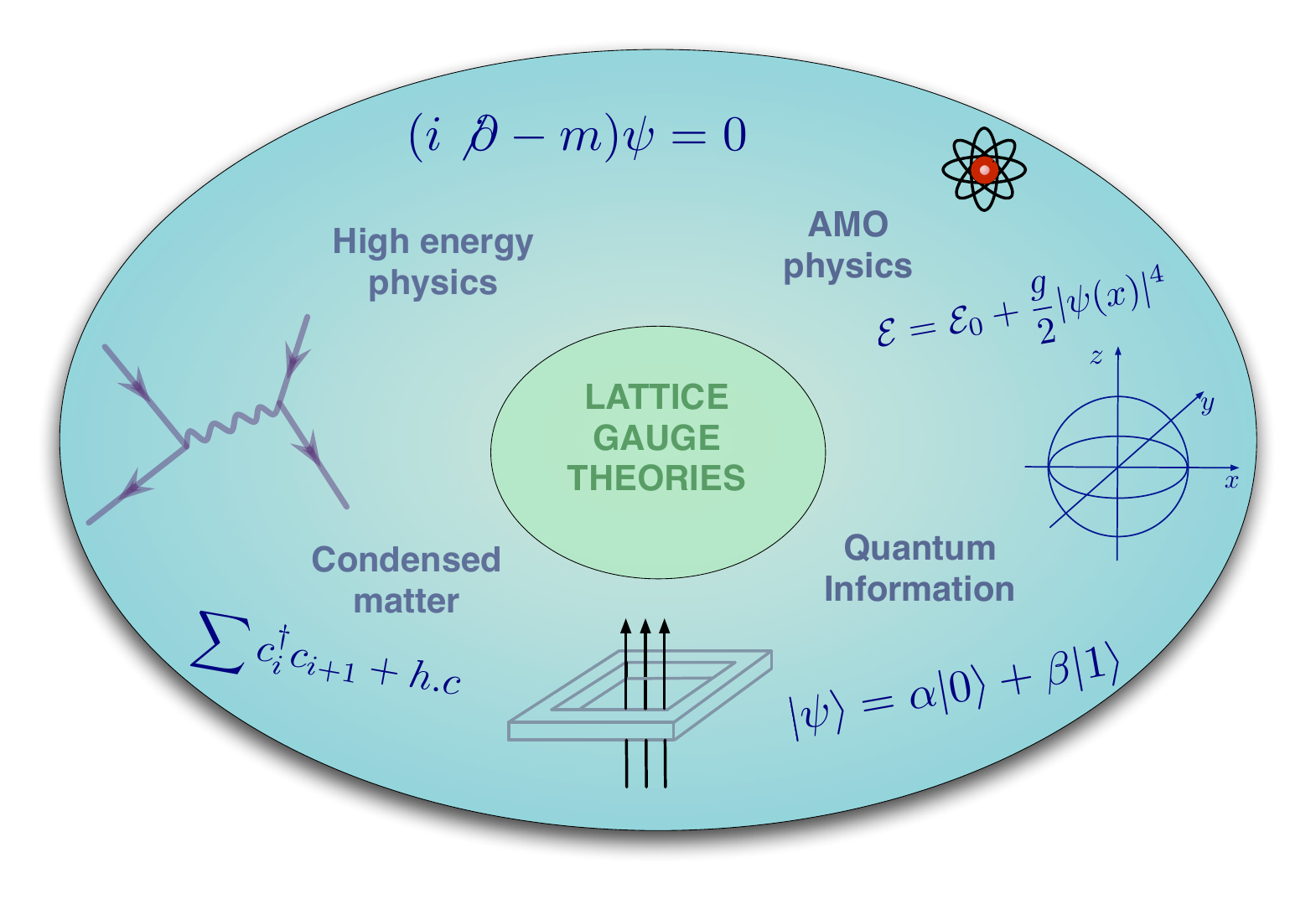}}
\caption{Quantum Lattice gauge theories lie in between different disciplines, as they describe systems in high-energy physics; atomic, molecular and optical (AMO) physics; condensed matter physics and quantum information science.}
\label{fig:multi}
\end{center}
\end{figure}

LGTs play a fundamental role in different fields, in particular in high-energy physics as the interactions between fundamental 
particles can be described by a gauge theory. We have already introduced QED, which describes the dynamics of electrons and photons. However, all fundamental particle interactions -- the Standard Model of high-energy physics- is a gauge theory~\cite{Buchmuller2006}. 
In condensed matter, gauge theories arise as low energy effective theories of many-body Hamiltonians~\cite{Lieb1993a,Lee2006,Lacroix2010} 
and finally, some schemes proposed to implement topological quantum computation are again described by gauge theories, like the Kitaev model~\cite{Nayak2008,Kitaev2003}. 
The recent efforts described in this review towards quantum simulators of LGT by the quantum information and the atomic, molecular and optical physics 
communities enlarged the number of fields of physics that encompass LGT in their object of study. Finally, the contributions to lattice calculations given by the 
computational physics and high-performance computing communities, make the study of LGT 
a truly multidisciplinary discipline at the center of different fields with a high number of 
interconnections as depicted in Fig.~\ref{fig:multi}.

~\\
This pedagogical review is composed as follows: in the next section we briefly recall the implications of the presence of global symmetries in quantum many-body systems; while in Section~\ref{IntroLGT} we introduce lattice gauge theories emphasizing the difference between global and local symmetries and between abelian and non abelian ones. In Section~\ref{TNLGT} we present the different approaches that have been 
introduced to study LGT via tensor network methods and the calculations achieved so far, and in Section~\ref{QsymLGT} we 
review the main ideas underlaying current proposals for quantum simulations of LGT along with some examples. Finally, in Section~\ref{conclu} our 
conclusions and expectations on the future evolution of this new exciting new field are presented.

\section{Global symmetries}
In any investigation approach one pursues, symmetries play a fundamental role: they provide a simplification 
of the problem and determine the properties of the system like the different phases of matter the system displays as described by the  
spontaneous symmetry breaking, a cornerstone of the theory of critical phenomena~\cite{VanWezel2007}. 
Indeed, in the presence of a conservation law -- for example if the total number 
of particles is fixed -- the problem might be drastically simplified: the first impact of a constant of motion is indeed the reduction of the number of the 
possible configuration the system can assume: with a single particle in a system of $N$ lattice sites,
the allowed configurations are drastically reduced and the dependence of the 
configuration space dimension changes from an exponential to a linear function of $N$. In general, one can show via a simple combinatorial argument that 
having a fixed number of particles $q$ implies that the Hilbert space dimension of the allowed states scales as $N!/(q! (N-q)!)$. 
That is, any state of the system at fixed number of particles $|\psi\rangle = \sum_{\vec \alpha \in \mathcal{Q}} \psi_{\alpha_1 \dots \alpha_N} |\alpha_1 \dots \alpha_N\rangle$ 
is such that the expectation value of the particle number $\hat{\mathcal{N}}=\sum_i \hat n_i$ (where the operator $\hat n$ counts the particle presents in each site) 
is constant and equal to $q$, that is, all states present in the set $\mathcal{Q}$ have the same $q$ number of particles. 
In particular, if the local states are labelled according to the particle number 
($\alpha =0$ labelling the state with no particles in the site, $\alpha =1$ with one particle and so on) the particle conservation translates 
in the constraint $\sum_i \alpha_i = q$. Equivalently, the symmetry 
implies that the Hamiltonian of the system commutes with the symmetry group generators, i.e, the corresponding matrix representations 
can be simultaneously diagonalized: the Hamiltonian can be written in block diagonal form, with each block labelled with the 
corresponding {\it good quantum number} $q$. In our example, each sector independently describes a fixed number of particles $q$, 
and thus one can work in a single sector.  

More formally, the conservation of the particle number is a global $U(1)$ symmetry,
that can be represented as a unitary matrix of dimension one or equivalently as a transformation on a circle:
in our example generated by the operator $\hat{\mathcal{N}}$. The infinitesimal transformation under which the Hamiltonian is invariant 
(which shall be unitary being a change of coordinates) is defined as $e^{i \hat{\mathcal{N}}\phi}$ (the Lie algebra generated by
$\hat{\mathcal{N}}$) where $\phi$ is a continuous parameter. The phase $\phi$ is the associated phase or order parameter 
in different paradigmatic scenarios as the Bose Einstein Condensate or in superconductors. 
Indeed, the Hamiltonian is invariant under the transformation $e^{i \hat{\mathcal{N}} \phi} H e^{- i \hat{\mathcal{N}} \phi} = H$ as 
it commutes with the number operator. Each sector with a given charge $q$ is transformed by $e^{i q \phi}$ and the corresponding eigenvectors are invariant 
under the transformation $e^{i \hat{\mathcal{N}} \phi} |\psi\rangle_q = e ^{i \sum_i \alpha_i \phi} |\psi\rangle_q = e ^{i q \phi } |\psi\rangle_q$ where 
the subscript is specified to stress that the each eigenvector lives in a sector with defined charge $q$. We note, for later comparison,  
that the invariant state is the full many-body one, the transformation acts on the full space and the conserved quantity is the total number of particles. 
In conclusion, in presence of a conserved scalar charge (being it an electromagnetic charge, a global spin, a number of particles etc.) 
the system is invariant under a global phase transformation proportional to the charge in each sector. 
In general, in presence of a symmetry transformation, the Noether theorem insures that there is a conserved quantity: for a $U(1)$ symmetry 
this is a scalar while, as we will see later on, it can be a more complex object for other symmetries~\cite{Goldstein1980}.

Given a global symmetry, we know from the theory of critical phenomena that spontaneous symmetry breaking might occur: there may be a phase where the value of the order parameter is non-zero and the system chooses one among the possible configurations, i.e., a single value of the order parameter~\cite{VanWezel2007}. 
For example, Ising ferromagnetism is a typical scenario of a $\mathbb Z_2$ broken symmetry, where the spins choose either 
to point north or south even though the system Hamiltonian is invariant under the spin flip operation. In the case of conservation of particle number, the symmetry that might 
be spontaneously broken is the $U(1)$ symmetry and the phase of the order parameter is what is fixed in the broken phase. The broken symmetry phase is a superfluid phase for bosons: a seminal work performed on cold atoms in optical lattices demonstrated the quantum phase transition between the Mott insulator and the superfluid phase, starting a fascinating series of experiments where quantum simulators have probed different physical phenomena in a controlled setup~\cite{Greiner2002,Bloch2012}.

\section{An introduction to lattice gauge theories}

\label{IntroLGT}

Since their introduction in 1974 by Wilson~\cite{Wilson74}, lattice gauge theories (LGTs) have found widespread application in the investigation of non-perturbative effects in gauge theories~\cite{Montvay1994,Creutz1997,DeGrand2006,Gattringer2010}. At the theoretical level, defining the theory on the lattice provides a natural regularization scheme, as the lattice spacing immediately introduces a cut-off in the theory. From the computational viewpoint, LGTs are amenable to classical simulation methods based on Montecarlo techniques. Differently from conventional lattice models found, e.g., in quantum statistical mechanics, LGTs are characterized by local symmetries and local conserved quantities, as a reflex of the invariance of the original theory under specific sets of transformations. 

In this Section, we present LGTs from a statistical mechanics point of view, which is well suited to describe their connections with  quantum information theory. While the most common formulation of LGTs is the Lagrangian one introduced by Wilson, which is an ideal setup for Monte Carlo simulations~\cite{Montvay1994,Creutz1997,DeGrand2006,Gattringer2010}, we discuss here the Hamiltonian formulation originally introduced by Kogut and Susskind~\cite{Kogut1975,Kogut1979}, using as a preferred framework the so-called quantum link models~\cite{Horn1981,Orland1990,Chandrasekharan1997}. Those models share many features with the conventional Wilson formulation -- most importantly, they keep gauge invariance exact on the lattice -- but are usually more amenable to both tensor network methods and quantum simulations thanks to the fact that their Hilbert space is finite-dimensional. In addition, they are intimately related to various problems in the context of frustrated quantum magnetism~\cite{Lacroix2010,Lee2006}, so that they provide a natural connection to condensed matter and statistical mechanism application of the aforementioned methods. We note that, for the case of discrete gauge groups (e.g., $\mathbb{Z}_N$), the quantum link and Wilson formulation actually coincide~\cite{wegner71,Kogut1979,Creutz1997}, and for specific group choices, other formulations of discrete gauge theories have also been introduced - see, e.g., Refs.~\cite{Mathur:2005cr,Anishetty:2010dq,Tagliacozzo2014,Zohar2015}

We will start our discussion reviewing the role of $U(1)$ global symmetries in fermionic systems, and discussing the case of a static background gauge field. Then, we will show how promoting a symmetry from global to local requires the introduction of new dynamical variables, the {\it dynamical gauge fields}. While the treatment will deal with $U(1)$ symmetries, we will present extensions to non-Abelian symmetries at the end of the section, where we will also comment on the connections to Wilson's formulation of LGTs and to gauge theories in the continuum.

\subsection{Free fermions on a lattice: $U(1)$ global symmetry}

\begin{figure}
\begin{center}
\begin{minipage}{60mm}
\resizebox*{6cm}{!}{\includegraphics{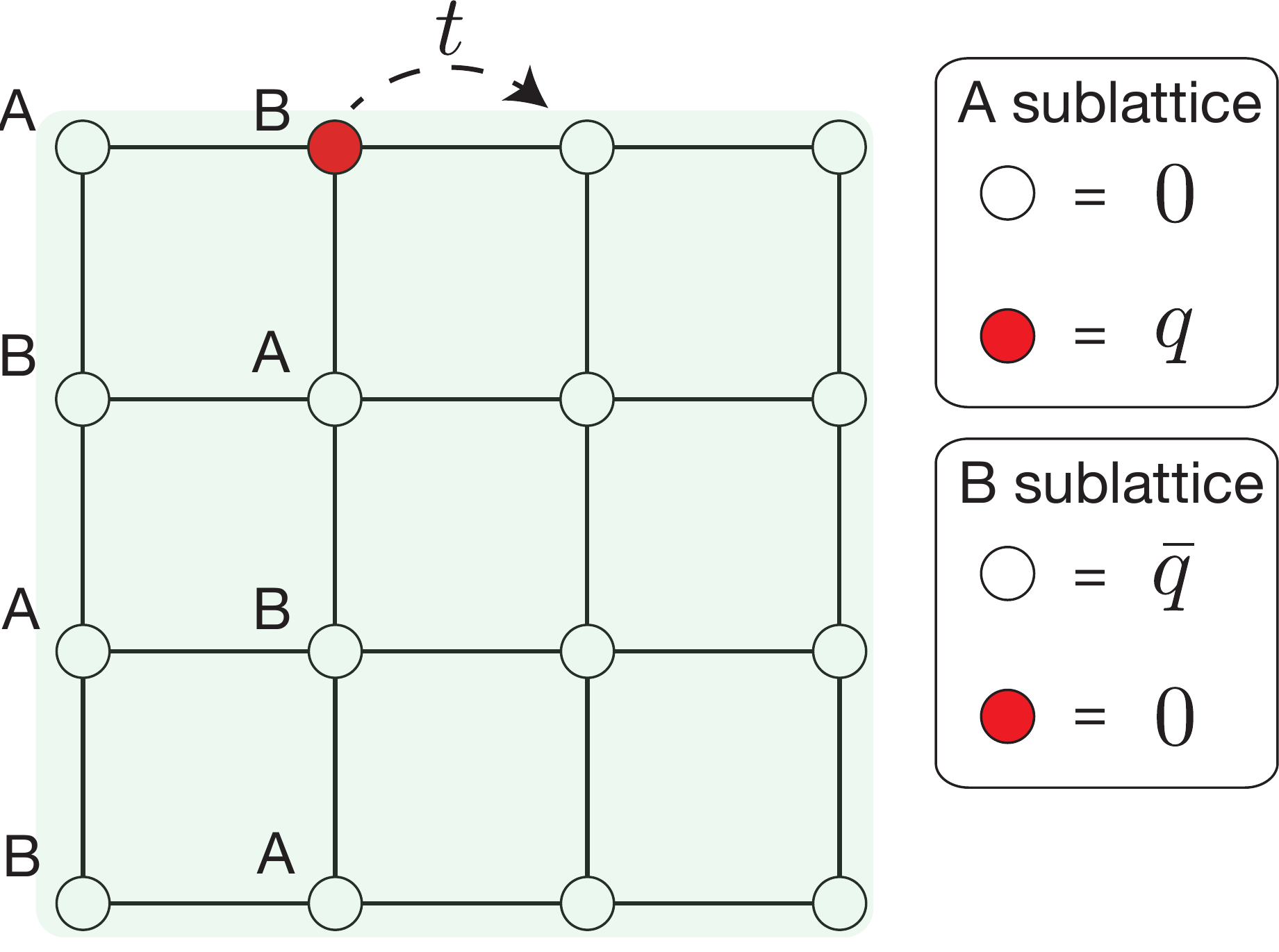}}\hspace{6pt}
\caption{Free fermions hopping on a 2d lattice as described by Eq.~\eqref{FF_ham}. The number of particles is conserved over the entire lattice (shaded area) as a consequence of a global U(1) symmetry.} \label{fig:freefermions}
\end{minipage}
\end{center}
\end{figure}

The simplest example of a global symmetry in many-body systems is probably number conservation. Let us consider fermions hopping on a two-dimensional (2D) square lattice, whose dynamics is described by an Hamiltonian of the class of Eq.\eqref{ham}:
\begin{equation}\label{FF_ham}
H = - t \sum_{<i, j>} (c^\dagger_{i}c_j+c^\dagger_{j}c_i) + m \sum_i (-1)^i n_i
\end{equation}
where $c^\dagger_j \, (c_j)$ are fermionic creation (annihilation) operators at the site $j$ of the lattice, $n_i = c^\dagger_ic_i$, $t$ is the tunneling matrix element, $m$ the mass of the particles, and the sum is over all nearest-neighbors. 
The factor $(-1)^i$ realizes the so-called {\it staggered} (or Kogut-Susskind) fermions~\cite{Kogut1975,Kogut1979} (see Fig.~\ref{fig:freefermions}), which represent particles on one sub lattice ($+1$, sub lattice 'A') and antiparticles on the other ($-1$, sub lattice 'B'). Since the dynamics only involves fermions hopping on the lattice, the total number of fermions is conserved. As shown in the previous Section, defining the total number operator as $\hat{\mathcal{N}} = \sum_i n_i$, one has that the Hamiltonian and $\hat{\mathcal{N}}$ commutes, that is:
\begin{equation}
[H, \hat{\mathcal{N}}] = 0.
\end{equation}
Thus, $\hat{\mathcal{N}}$ corresponds to a conserved charge of the symmetry transformation. In particular, if we apply the following transformation on our lattice fermions:
\begin{equation}
c_i \rightarrow e^{i\alpha}c_i, \quad c_i^\dagger \rightarrow e^{-i\alpha}c_i^\dagger, \quad \alpha\in \mathbb{R}
\end{equation}
the Hamiltonian is mapped into itself. In order to make things a bit more formal, we could define a transformation of the form:
\begin{equation}\label{trasnfU1}
c_i \rightarrow Vc_i, \quad c_i^\dagger \rightarrow c_i^\dagger V^\dagger,
\end{equation}
where the operators $V$ act on the fermion subspace, and are elements of some group $\mathcal{G}$: for the specific case defined above, the group is $U(1)$, whose elements can be parametrized as phases $e^{i\alpha}$ as they span a circle in the complex plane.
Then, we could say that $H$ is invariant under transformations of the $\mathcal{G}$ group, if $H\rightarrow H$ after applying the transformation in Eq.~\eqref{trasnfU1}.

In the realm of many-body systems, there are various classes of interactions on top of $H$ which either preserve, or violate, such $U(1)$ symmetry. All interactions which are composed by only number operators, as the mass terms in Eq.~\eqref{FF_ham} or density-density ones of the form $H\propto \sum_{(i,j)}n_in_j$, are invariant under $U(1)$ transformations, while superconducting terms of the form $H\propto \sum_{<i,j>}(a^\dagger_{i}a^\dagger_j+ a_j a_i)$ are not~\footnote{Indeed, the breaking of the $U(1)$ symmetry down to $\mathbb{Z}_2$, symmetry known as parity}. This implies that the presence or absence of a symmetry drastically affects the possible couplings in a lattice model: as we will see below, this feature is further emphasized in the case of gauge symmetries, and represents a key element (and challenge) for designing quantum simulators for lattice gauge theories.

\subsection{Particles coupled to a dynamical gauge field: U(1) gauge symmetry}

Hereafter we show what happens if we try to promote a symmetry from global to local: in the previous example, 
the corresponding local transformation is:
\begin{equation}\label{gauge1}
c_i \rightarrow e^{i\alpha_i}c_i, \quad c_i^\dagger \rightarrow e^{-i\alpha_i}c_i^\dagger, \quad \alpha\in \mathbb{R}
\end{equation}
that is, the phase factor $\alpha$ now depends on the lattice index $i$. If we perform this transformation only at one site, then the Hamiltonian term involving the site $i$ and $j$ transforms as:
\begin{equation}
 - t  (c^\dagger_{i}c_j+c^\dagger_{j}c_i) \rightarrow  - t \sum_{<i, j>} (e^{i\alpha_i}c^\dagger_{i}c_j+e^{-i\alpha_i} c^\dagger_{j}c_i).
\end{equation}
We see immediately that this term is not invariant under Eq.~\eqref{gauge1}: that is, the system does not have a local U(1) symmetry. At an intuitive level, this can be understood by looking at conserved charges: the tunneling matrix element does not conserve the number of particles at the local level (as one site can turn from occupied to empty), but only globally in the whole system. 

In order to realize a local symmetry, a new set of operators has to be introduced. In the standard formulation of Wilson's lattice gauge theories, one introduces elements of a gauge group at each link: for continuos groups, these operators usually span an infinite-dimensional Hilbert space (which for the $U(1)$ case is a direct reflection of the fact that the local Hilbert space of photons is not bound). This approach is reviewed in various books in a formal and elegant way~\cite{Montvay1994,Creutz1997,DeGrand2006,Gattringer2010}. Here, we focus instead on the construction of gauge theories which have discrete, finite-dimensional Hilbert spaces on the links. These models, known as quantum link models (or gauge magnets), have been introduced in the high energy context in a series of papers by Horn~\cite{Horn1981}, Rohrlich and Orland~\cite{Orland1990}, and Brower, Chandrasekharan and Wiese~\cite{Chandrasekharan1997,Brower1999}. In parallel, some of those models have also been discussed in the condensed matter context in the Abelian case~\cite{Lacroix2010}, and have close ties to some exactly soluble models in classical statistical mechanics~\cite{baxterbook}. While they differ in some aspects to the conventional Wilson's formulation of lattice gauge theories (see, e.g., the discussion in Ref.~\cite{Wiese:2013kk}), they keep gauge invariance exact, which is the basic feature we are interested in here. Moreover, they are also naturally formulated in the statistical mechanic language of Hubbard models~\cite{Brower1999}, which makes them an ideal framework to cast several concepts of lattice gauge theories in both tensor network and quantum simulation contexts.

\begin{figure}
\begin{center}
\begin{minipage}{160mm}
\resizebox*{16cm}{!}{\includegraphics{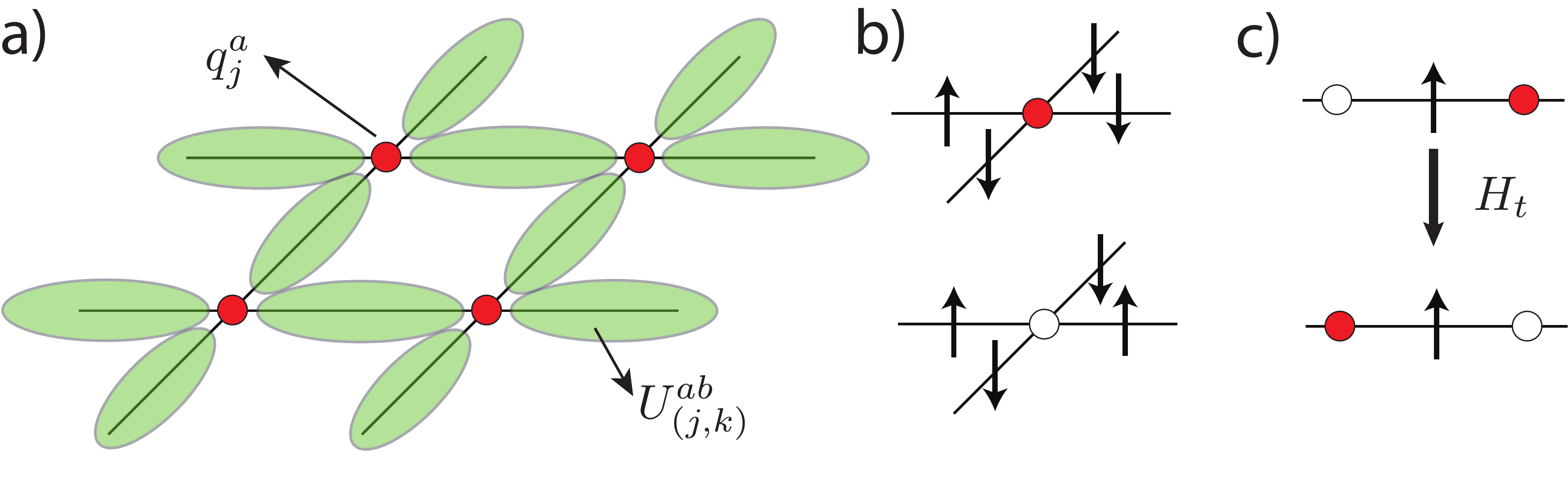}}\hspace{6pt}
\caption{Schematic illustration of a U(1) lattice gauge theory. Panel a) configuration space of a lattice gauge theory, with matter fields of charge $q^a_j$ on the vertices, and gauge fields $U^{ab}_{ij}$ on the bonds. Panel b) examples of some gauge invariant states satisfying the Gauss's law in Eq.~\eqref{GaussU1} at odd sites.  Panel c) schematic link dynamics driven by the term in Eq.~\eqref{Htt}.} \label{LGTFigure}
\end{minipage}
\end{center}
\end{figure}

We start our construction by introducing spin variables $\bar{S}_{(i,j)}$ on each bond $(i,j)$ which serve as {\it gauge fields}~\footnote{For the sake of compactness, we assume that, for horizontal bonds, the index $(i,j)$ is always ordered from left to right, while for vertical bonds, it is order top to bottom}. For simplicity, we employ spins in the $S=1/2$ representation - the construction extends trivially to higher spins as well~\cite{Banerjee2012}. Before explicitly constructing the Hamiltonian of the system, it is instructive to discuss its Hilbert space. Initially, the Hilbert space is naturally the tensor product of the fermions $\mathcal{H}_F$ and the gauge fields $\mathcal{H}_S$, separately, that is: $\mathcal{H} = \mathcal{H}_F \otimes \mathcal{H}_S$. Within this space, we need to identify a subspace defined by a set of local conserved charges. In order to do that, we analyze the Hilbert space of a single vertex, which include the fermionic mode at the center, and the four spins residing on the bonds including the site itself (see Fig.~\ref{LGTFigure}a).

Working on the diagonal basis of the fermionic number operator and the spin $z$-direction, it is possible to define the following operator at each vertex:
\begin{equation}\label{GaussU1}
G_i = n_i+\sum_{(i,j)\in +} (-1)^{ij}S^z_{(i,j)} + \frac{(-1)^i +1}{2}
\end{equation}
which takes integer values, and thus, is a good candidate for being the conserved charge of a $U(1)$ symmetry. The notation $(-1)^{ij}$ indicates a $+1$ for spins at the left and top of each vertex, and a $-1$ for spins at the right and bottom. We can then postulate that this operator works as a generator of a $U(1)$ transformation, and thus defines what is known as the gauge invariant Hilbert space $\mathcal{H}_{G}$, which is defined as the space spanned by all physical states $|\Psi\rangle$ such as:
\begin{equation}
\mathcal{H}_{G} = \{|\Psi\rangle \; | \; G_i|\Psi\rangle =0 \; \forall i\},
\end{equation}
where the condition $G_i|\Psi\rangle =0$ is the so-called {\it Gauss's law}. All states which do not satisfy Gauss's law are known are {\it unphysical states}, and form the so called gauge-variant Hilbert space:
\begin{equation}
\mathcal{H}_{\textrm{unphys}} =\mathcal{H}  \setminus \mathcal{H}_G=\{|\Psi\rangle \; | \;\exists i  \; \textrm{such that} \; G_i|\Psi\rangle \neq0 \}.
\end{equation} 
At the single link level, the local Hilbert space is illustrated in Fig.~\ref{LGTFigure}b for a vertex on the odd sublattice ($(-1)^i=-1$). Out of the possible $2^4 * 2 = 32$ possible configurations of a single vertex, only 10 are allowed by the Gauss's law: in case there is a particle sitting at the vertex, the sum of the total spins on the bonds should give $-1$, leaving 4 possible configurations, while in case there is no particle, 6 spin configurations are allowed. 

As seen before, the presence of the Gauss's law in this statistical mechanism models allows a direct interpretation of the lattice fields to conventional electrodynamics, where the Gauss's law reads $\rho = \nabla \vec{E}$. At each vertex, the charge operator $\rho$ corresponds to $n_i + ((-1)^i -1)/2$ because of the nature of the staggered fermions: on the A-sublattice, the presence of a fermion corresponds to a particle, so there is one charge unit, while on the B sublattice, a hole corresponds to an antiparticle, thus giving minus one charge unit. The relation between spin and electric fields also follows a staggered pattern, that is $E_{(i,j)} = S^z_{(i, j)} * (-1)^i$. In such  way one can interpret the sum of spins in the generator $G_i$ as a discrete derivative considering the horizontal bonds $(k,i), (i,j)$ from left to right, respectively, the sum $S^z_{(k,i)} + S^z_{(i,j)}$ maps onto the difference $E_{(k,i)} - E_{(i, j)}$, which is nothing but the lattice approximation of the divergence $\nabla \vec{E}$ along the horizontal direction (the same applies to the vertical one).

The formulation of the local Hilbert space is the starting point to understand which Hamiltonian dynamics is compatible with the Gauss's law, that is, $[H, G_i]=0\;\forall i$. First, as in the presence of global symmetries, all terms involving solely operators  diagonal in the local basis (e.g. $n_i, S^{z}_{(i,j)}$) are gauge invariant, and are thus allowed. We will come back to those at the end of the section, and discuss their relevance in various scenarios. The other terms which are not diagonal could involve {\it i)} the matter fields, {\it ii)} the gauge fields, or {\it iii)} both matter and gauge fields. 

The first class of terms is clearly not allowed: acting with whatever combination of $a_i,a_j, ...$ operators will immediately break the gauge constraint at the site $i, j, ...$. In particular, this combination of operators will introduce particles in the system without re-arranging the gauge fields. This reflects the fact that, in a gauge theory, particles and antiparticles cannot be created without exciting the gauge fields as well: this is analogous to the fact that in QED, the presence of two charges is always accompanied by an electric flux string connecting the two. 

On the other hand, terms of the second class, acting solely on gauge fields are allowed. 
An example of those is the so-called ring-exchange term:
\begin{equation}
H_{\square} = -J \sum_{\square} (S^{+}_1S^+_{2}S^-_3S^-_4 + \textrm{h.c.})
\end{equation}
which has been widely discussed in the condensed matter context~\cite{Lacroix2010}, and represents the equivalent of a magnetic field energy term in quantum electrodynamics. Note that here we adopt a slightly different notation with respect to the more conventional condensed matter one, which can be obtained using the transformation $\tilde{S}^z_i\rightarrow (-1)^iS^z$. Here, the sum involves all plaquettes in the square lattice, and the indices $1,2,3,4$ denote the bonds in each plaquette in clockwise order. It is easy to show that this is the minimal interaction involving the gauge fields only, as one-, two-, and three-body terms off-diagonal in the $z$-basis would violate gauge invariance at least at 2 vertices. We note that the ring-exchange term is only the simplest version of gauge field interactions: indeed, there exist terms involving larger, closed loops in the lattice which are also gauge invariant. 

The last class of terms involves interactions between the matter and the gauge fields. The most natural term in this class describes the matter field tunneling dynamics coupled to the gauge fields, and reads
\begin{equation}\label{Htt}
H_{t} = -t \sum_{<i,j>} (c^\dagger_i S^{+}_{(i, j)}c_{j} + c^\dagger_j S^{-}_{(i, j)}c_{i} ).
\end{equation}
In this term the tunneling dynamics of a fermion is always accompanied by a spin flip on the bond between the two matter sites, in order to be consistent with the Gauss's law. The action of this term on a cartoon state is schematically illustrated in Fig.~\ref{LGTFigure}c, and differs fundamentally from the conventional dynamics described by Eq.~\eqref{FF_ham}. In particular, these terms are not quadratic in any basis, so this implies that interactions among gauge and matter fields are always present.

The total Hamiltonian of staggered fermions coupled to a $U(1)$ gauge field in (2+1)-d then reads:
\begin{equation}\label{QLM_U1}
H_{U(1)} = H_{t}  + H_{\square} + m \sum_i (-1)^i n_i + V \sum_{\square} (\prod_{j\in\square} S^z_j) + \kappa\sum_{j} (S^z_j)^2
\end{equation}
where we have also added a staggered mass $m$ for the fermions, an additional potential energy $V$ for flippable plaquettes, and the equivalent of an electric field energy density $\kappa$ (which is non-trivial for spin representations $S>1/2$). We note that in the one-dimensional limit, where the magnetic field term $H_\square$ disappears, a discrete lattice version of the Schwinger model (QED in (1+1)-d) is recovered. For practical purposes, it is useful to rewrite the spin operators as a function of bosons bilinears using the Schwinger representation, where~\cite{Banerjee2012}:
\begin{equation}\label{Sbosons}
S^+_j = a^\dagger_jb_j, \quad S^z_{j} = \frac{n^a_j-n^b_j}{2}, \quad [a_j, a^\dagger_i] =  [b_j, b^\dagger_i] = \delta_{ij}.
\end{equation}
These bosons are called {\it rishon} in the quantum link formulation: such link decomposition will be important for the formulation of gauge invariant tensor network ansatze, and leads to natural generalizations to non-Abelian symmetries, as discussed below. 

In the next subsections, after providing a comparison between static and dynamical gauge fields, we review some useful cases of interest for the next sections, namely quenched and non-Abelian lattice gauge theories.

\begin{figure}
\begin{center}
\begin{minipage}{160mm}
\resizebox*{8cm}{!}{\includegraphics{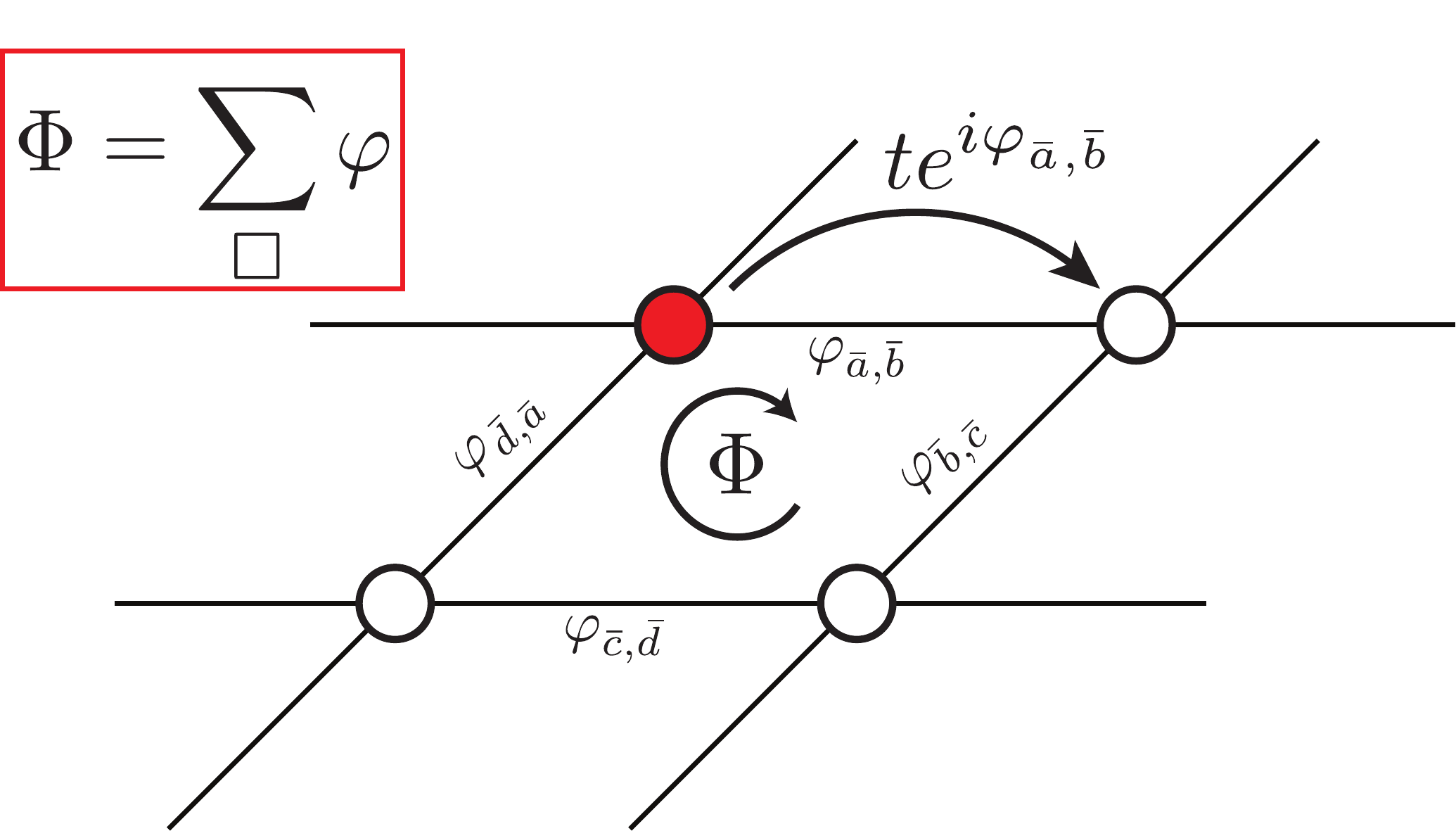}}\hspace{6pt}
\resizebox*{8cm}{!}{\includegraphics{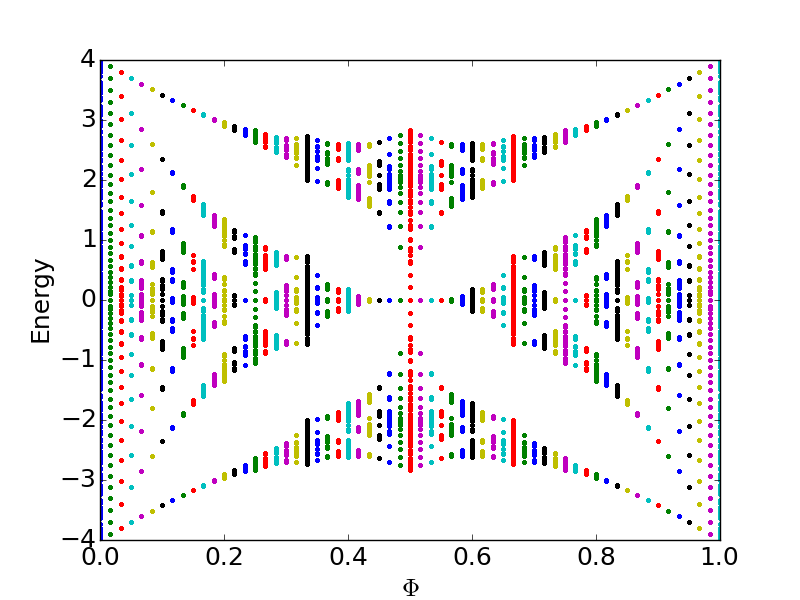}}\hspace{6pt}
\caption{Schematic illustration of the Hofstadter model (a). Particles move on a square lattice, where the tunneling matrix element takes phases $\varphi_{\bar{a},\bar{b}}$ which depend on the bond between each pair of sites (sites in the plaquette are labeled as $(a,b,c,d)$ from top-left in clockwise order). The total flux per plaquette $\Phi$ corresponds to the total phase accumulated by the wave-function of a single particle which moves around the plaquette itself. The spectrum of the system is described by the Hoftstadter butterly for a lattice of 60x20 sites.} \label{HofFigure}
\end{minipage}
\end{center}
\end{figure}

\subsubsection{A comparison to particles coupled to a background gauge field: the Hofstadter model}

In condensed matter and cold atom systems, there is another type of gauge fields,  the so called {\it background} (or static) gauge fields, which has attracted a lot of attention - in particular in the context of topological quantum matter~\cite{BernevigBook}. An archetype example is the Harper-Hofstadter Hamiltonian:
\begin{equation} 
H_{HH} = - t \sum_{<i,j>} (e^{i\phi_{ij}}b^\dagger_ib_j+\textrm{h.c.})
\end{equation}
describing non-interacting fermions in a 2d square lattice, whose hopping is associated with a phase factor $\phi_{ij}$ in such a way that, when tunneling over a single plaquette, the fermion wave functions picks up a phase factor $\Phi = \sum_{(ij)\in\square}\phi_{ij}$. This quantity can be interpreted as a flux of an effective magnetic field piercing each plaquette, and is instrumental in determining the phase structure of the spectrum, which displays the celebrated Hofstadter butterly structure (see Fig.~\ref{HofFigure}).

However, those static gauge fields, being classical variables, are fundamentally different from the dynamical gauge fields we are interested in. The main difference is that there is no back-action of the matter fields onto the gauge fields, since the latter have no quantum dynamics. At a symmetry level, this implies that there are no conserved local charges, so the gauge symmetry as described by the Gauss's law in Eq.~\eqref{GaussU1} is absent.  

\subsubsection{Quenched gauge theories: quantum Ice on the square lattice}

Contrary to the case of classical gauge fields, gauge theories can be defined even in the absence of any matter: this approximation is usually called {\it quenched}, as the matter fields are frozen in given configurations. At the computational level, these theories are much easier to handle via Monte Carlo simulations, as the sign problem is usually only due to the fermionic matter. Another interesting aspect is that some of these quenched gauge theories have actually attracted quite some interest in the condensed matter realm, where they are usually referred to as quantum dimer or quantum loop models (with the difference being the static  configuration of background charges). 

A paradigmatic example of a quenched quantum link model is the so called (2+1)-d quantum Ice model, whose Hamiltonian reads:
\begin{equation}\label{QLM_U1}
H_{QI} = H_{\square} +  V \sum_{\square} (\prod_{j\in\square} S^z_j) 
\end{equation}
and acts on a Hilbert space which is defined by the so-called {\it Ice rule}. The model displays a series of magnetic phases with classical ordering in the regime $V>1, V\lesssim -0.33$, and a plaquette resonating-valence-bond phase in the intermediate regime~\cite{Shannon:2004nr,Banerjee:2013gf}. Moreover, the same Hamiltonian can describe a quantum dimer model, in the gauge sector where static charges are introduced on a lattice bipartition (similar to the vacuum state of staggered fermions).

\subsection{Non-Abelian lattice gauge theories}

Finally, we discuss briefly the extension to non-Abelian gauge symmetries, symmetries whose generators do not commute. The discussion becomes pretty soon fairly technical, however for completeness we report the fundamentals here below; for a comprehensive review on the topic, see Refs.~\cite{Brower1999,Wiese:2013kk,Zohar2015a}. While, even for discrete systems, many formulations of non-Abelian gauge theories have been introduced, we focus here on the rishon formulation of quantum link models, which is sufficiently general to cover $U(N)$ and $SU(N)$ models in a compact form. The index $N$ related to the non-Abelian group denotes the presence of different `color' degrees of freedom in the theory. In our discussion, we will mostly follow the original presentation of Ref.~\cite{Brower1999}.

For simplicity, let us assume a one dimensional geometry, and discuss the $SU(N)$ case: at each link, instead of a single component field, we introduce a pair of fermionic {\it rishon} operators, $c^{\alpha}_{R; j}, c^{\beta}_{L; j+1}, \alpha, \beta \in [1, N]$, where $\alpha, \beta$ are color indices.
The non-Abelian gauge field is then built up as a fermion bilinear as follows:
\begin{equation}
U^{\alpha\beta}_{(j, j+1)} = c^{\alpha}_{R; j} c^{\beta\dagger}_{L; j+1}
\end{equation}
while at the vertices, we introduce fermionic operators $\psi^\alpha_{j}$. The fundamental building block of this construction consists of a single matter field site $j$, and the two fermionic link sites close by, namely $[L; (j-1,j)]$ and $[R, (j, j+1)]$. For $SU(N)$, we have $N^2-1$ generators $G_j^\gamma$, given by:
\begin{equation}
G^\gamma_j = \psi^{\alpha\dagger}_j\lambda^{\gamma}_{\alpha\beta}\psi^{\beta}_j + L^\gamma_{j}+R^{\gamma}_j
\end{equation}
where $\lambda^\gamma$ are the SU(N) generalized Gell-Mann matrices, and the $L, R$ are the non-Abelian flux operators, and are functions of the rishon operators as:
\begin{equation}
L^\gamma_j = c^{\alpha\dagger}_{L;j}\lambda^\gamma_{\alpha\beta}c^\beta_{L,j}, \quad R^\gamma_j = c^{\alpha\dagger}_{R;j}\lambda^\gamma_{\alpha\beta}c^\beta_{R,j},
\end{equation}
The generators define the Hilbert space according to the Gauss's laws $G^\gamma|\Psi\rangle=0$, exactly as in the Abelian case before. The gauge-invariant Hamiltonian dynamics is described by~\cite{Banerjee2013}:
\begin{eqnarray}
H_{SU(N)} &=& -t\sum_{j} [(\psi^{\alpha\dagger}c^{\alpha}_{R; j}) (c^{\beta\dagger}_{L, j+1}\psi^\beta_{j+1}) + \textrm{h.c.}] + m \sum_j (-1)^j\psi^{\alpha\dagger}_j\psi^\alpha_j +\\
&+& \frac{g^2}{2} \sum_j [L^\alpha_{j}L^\alpha_{j} + R^\alpha_{j}R^\alpha_{j}] +\epsilon\sum_j (\prod_{\kappa=1}^N c^{\kappa\dagger}_{R; j}c^\kappa_{L; j+1}+ \textrm{h.c.})
\end{eqnarray}
where $\textrm{h.c.}$ stands for an operator Hermitian conjugate of the previous one; the first term in the second line is a non-Abelian electric field; and the last term represents an $N$-particle rishon tunneling over the bond (for the case $\epsilon =0$, the symmetry is extended to $U(N)$), which implies that the number of rishon per bonds has to be $N$.

\subsection{Connections to Wilson lattice gauge theories, the continuum limit, and other discrete formulations of gauge theories}

While we have focused up to now on the discrete formulation of lattice gauge theories, the Wilson formulation, which introduces parallel transporters as link variables, has been the widespread framework for numerical studies in high energy physics. A valuable introduction of the Wilson approach in the Hamiltonian formulation can be found in Kogut's review article~\cite{Kogut1979}. Very often, one is ultimately interested in taking the continuum limit of a lattice field theory: in this respect, the Wilson and quantum link formulation differ substantially, as one relies on taking the limit of vanishing lattice spacing~\cite{Kogut1979}, while the other usually employs a mechanism known as dimensional reduction~\cite{Brower1999,Beard1998775}. For a detailed discussion of this aspect, we refer the reader to Ref.~\cite{Wiese:2013kk}.

Finally, we point out that discrete gauge theories can also be defined using different schemes than the quantum link model, such as the prepotential formalism~\cite{Mathur:2005cr,Anishetty:2010dq}, or truncation schemes based on the group representations~\cite{Tagliacozzo2014,Zohar2015}. Those approaches have also found application in the context of quantum and classical simulation of gauge theories: we will discuss some specific example in the next sections.

\section{Tensor network simulation of lattice gauge theories}
\label{TNLGT}

We have seen in the previous sections that lattice gauge theories are by definition defined over many different local sites, each of them representing a quantum degree of freedom, either bosonic or fermionic. 
The system wave function lives in the Hilbert space formed by the tensor product of all the local ones $|\alpha_1\rangle \otimes |\alpha_2\rangle \otimes \dots |\alpha_N\rangle$,
giving the amplitude of probability of each possible system configurations: thus, 
a generic state on the lattice is described by the exponentially increasing N-rank tensor 
\begin{equation}
| \Psi \rangle = \sum_{\{\alpha_i\}} \psi_{\alpha_1 \alpha_2 \dots \alpha_N}  |  \alpha_1 \alpha_2 \dots \alpha_N\rangle,
\end{equation} 
where $\{\alpha_i \} = \alpha_1, \alpha_2, \dots \alpha_N$ and $\mathcal{P}_{\alpha_1 \alpha_2 \dots \alpha_N} =|\psi_{\alpha_1 \alpha_2 \dots \alpha_N}|^2$ gives the probability to find the system in the configuration $\alpha_1 \alpha_2 \dots \alpha_N$ . 
Hereafter, unless otherwise specified, we will label the local Hilbert space via the local occupation number, i.e. $|\alpha \rangle = (b^\dagger)^k | 0 \rangle$.  For lattice gauge theories in the quantum link representation all local basis are finite, however in general 
for bosonic operators it is possible to truncate the local Hilbert space at some large value such that the physics of the system is not influenced by this approximation (or, at least, errors introduced by the truncation scheme can be controlled).
In conclusion, the local Hilbert space has some finite dimension $d$, which might be also site dependent even though here for simplicity we consider it uniform throughout the whole lattice. 

The  TN approach stems from two considerations: i) the tensor $ \psi_{\alpha_1 \alpha_2 \dots \alpha_N}$ can be viewed as a simple generalization of a matrix 
and thus, it might be possible to compress its information content in a compact form and ii) the tensor structure of the Hilbert space hints to the fact that $\psi_{\alpha_1 \alpha_2 \dots \alpha_N} $ might not be (typically) 
a fully general rank-N tensor but might have some properties which allow for an efficient representation. Building on this and other more formal considerations related on the amount of quantum correlations that are typically present in 
many-body systems~\cite{Eisert2010}, one can build classes of states that are defined via their tensor structure representing the state~\cite{Schollwock2011,Lubasch2014,Verstraete2004,Kliesch2014b,Tagliacozzo2009,Evenbly2012,Gerster2014b,Vidal2007}. 
The most natural one is the rank-N tensor introduced above, however one can propose many others 
which are equivalent unless approximations are introduced. Notice that one fixes the tensor structure, but not the dimension of the tensor indices nor 
the elements of each tensor: they are the variables used to 
build the most efficient faithful representation of the system under study, given the desired precision or amount of resources available. 

The simplest example of this approach is actually used in many different fields, and it is commonly known as mean field approximation. Indeed, within 
the mean field approximation one assumes that correlations shall be neglected, 
that is, the many-body wave function is the product of $N$ independent single body wave functions. This assumption corresponds to impose the tensor structure of the wave function as follows:
\begin{equation}
 \psi^\textrm{MF}_{\alpha_1 \alpha_2 \dots \alpha_N}  = \psi_{\alpha_1} \cdot \psi_{\alpha_2} \dots \psi_{\alpha_N}; 
 \label{meanfield}
\end{equation}
that is, one assumes that the original n-rank tensor can be decomposed exactly as $N$ rank-1 tensors.
Thus, the object of study has been simplified from an exponential to a linear problem in the number of lattice sites $N$.  An additional assumption results in an additional simplification of the problem: translational invariance implies that the 
$N$ tensors are equal, recasting the original problem into a problem independent of the system size $N$. 
Indeed, the number of degrees of freedom 
(free parameters in the tensors) is $d^N$ in the original problem, $N d$ in the mean field scenario, and only $d$ in the translationally invariant one. It is then not surprising that it is possible to find efficient algorithms or even analytical solutions to 
very complex problems in the mean field approximation. However, the drawback of this powerful approach is the fact that the introduction of Eq.~\eqref{meanfield} is in general unjustified and uncontrolled. Indeed, independent 
checks of the mean field approximation are always needed, either via comparison with experimental results or by other means. 

The TN ansatzes can be seen as generalization of the mean field approach and allow to introduce a controlled approximation which interpolates between the mean field solution and the exact one. In particular, for one dimensional systems, the class of Matrix Product States 
plays this role and it is defined as:
\begin{equation}
| \Psi \rangle_{\textrm{MPS}} = \sum_{\{\alpha_i, \beta_i\}} A_{\alpha_1}^{\beta_1}  A_{\alpha_2}^{\beta_1\beta_2}  A_{\alpha_3}^{\beta_2\beta_3}  \dots A_{\alpha_N}^{\beta_{N-1}}  |  \alpha_1 \alpha_2 \dots \alpha_N\rangle,
\label{MPS}
\end{equation}  
where the $A$s are rank-3 tensors (rank-2 at the boundaries, the first and last sites) connecting the physical indexes $\alpha$s with some auxiliary ones $\beta$s, whose dimension is $m$, i.e. $\beta_i= 1, \dots m$. 
To understand the role of these auxiliary indexes it is useful to make a simple example, for the case of two sites, that is to express as a MPS a two body wave function $\psi_{\alpha_1,\alpha_2}$: indeed in this 
case it is straightforward to identify the rank-2 tensor $\psi_{\alpha_1,\alpha_2}$ with a matrix, being it an object with two indexes. Thus, we can apply the whole linear algebra tools available to manipulate matrices, 
and in particular the Singular Value Decomposition (SVD), which allows to decompose any matrix $M$ as a product of two unitary matrices $S$ and $D$ and a diagonal one $V$, containing the singular values, such that 
$M= S V D$. If we apply the SVD to the two body wave function and write explicitly the indexes, we obtain:
\begin{equation}
\psi_{\alpha_1,\alpha_2} = \sum_{\beta_1,\beta_1'} S_{\alpha_1,\beta_1} V_{\beta_1,\beta_1'} D_{\beta_1',\alpha_2} \equiv  \sum_{\beta_1}A_{\alpha_1}^{\beta_1}  A_{\alpha_2}^{\beta_1},
\end{equation} 
where the diagonal matrix $V$ has been absorbed in one of the two $A$ matrices, i.e. $A_{\alpha_2}^{\beta_1}= \sum_{\beta_1'}  V_{\beta_1,\beta_1'} D_{\beta_1',\alpha_2} $. Notice that the dimensions of the indexes are all equal and equal to the Hilbert space local dimension $d$. 
However, if some singular values $\lambda_i$ are small enough, one could reduce the dimension of the auxiliary index $\beta_1$ to some value $m < d$ such that 
$ |\lambda_i| < \epsilon, \forall i > m$, for some arbitrary precision $\epsilon$.  The MPS state is 
nothing more than a generalization of this 
scenario for the case of $N$ local sites or a rank-N tensor. The only difference is that the dimension of the auxiliary dimension is site dependent as it increases moving through the chain. Indeed, to perform the SVD of a rank-N tensor one should first regroup the $N$ indexes in two subgroups: This is always possible and corresponds  to a reshuffling and 
relabeling of the tensor entries, e.g. $\psi_{\alpha_1 \alpha_2 \dots \alpha_N}  = \psi_{\{\alpha1\},\{ \alpha_2, \cdots, \alpha_N\}}= M_{i,j}$ where $i= \alpha_1 = 1, \dots d$ and 
$j = \{ \alpha_2, \cdots, \alpha_N\} = 1, \dots d^{N-1}$ is a rectangular matrix. Repeating the process regrouping a decreasing number of sites results in a MPS structure. Notice that, as an MPS with auxiliary dimension $m=1$ is a mean field ansatz given in Eq.~\eqref{meanfield}, the class of MPS interpolates 
between the mean field approximation and the exact (inefficient) description obtained for $m=2^N$.

\begin{figure}
\begin{center}
\begin{minipage}{160mm}
\resizebox*{16cm}{!}{\includegraphics{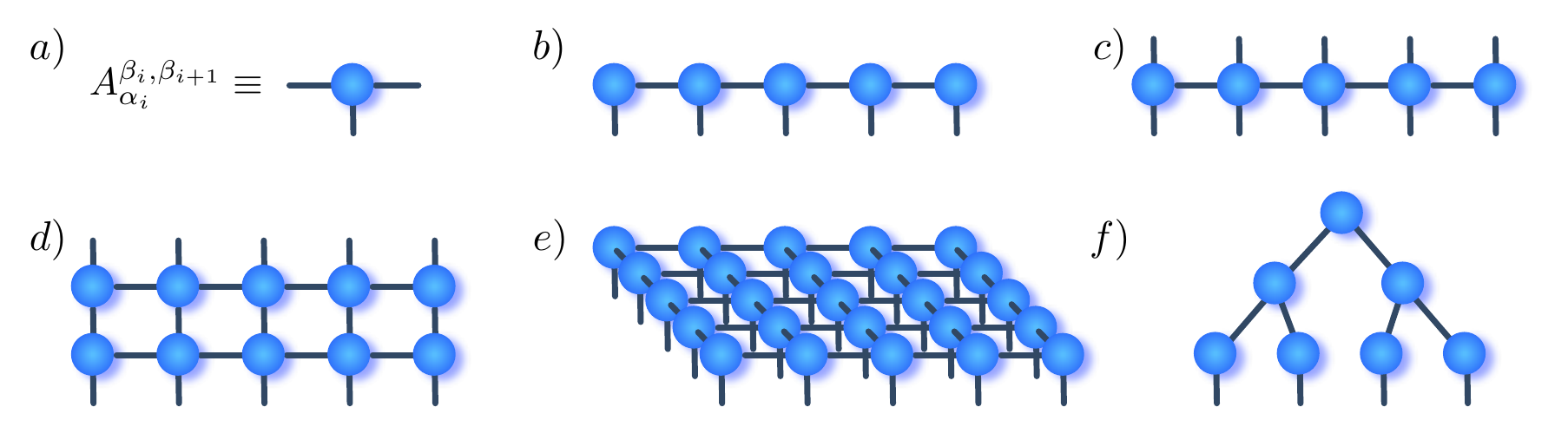}}\hspace{6pt}
\caption{Tensor network ansatzes: a) Pictorial representation of rank-3 tensors. b) Matrix Product State tensor network representation. c) TN representation of operators, a Matrix Product Operator. d) Locally purified TN: positive operators can be represented as $\rho= X^\dagger X$, where each $X$ operator is in MPO form. e) Two dimensional Projected Entangles Pair States TN representation. f) Tree Tensor Network hierarchical structure.   
} \label{TNansatz}
\end{minipage}
\end{center}
\end{figure}

The above described procedure allows to build an exact MPS representation of a rank-N tensor, however this is only a formal construction. Indeed, in practice performing the SVD of the rank-N tensor becomes hardly manageable for $N$ between ten and twenty. 
Thus, the starting point of MPS algorithms --and of all TN based ones-- is to assume that a faithful TN description of the state is possible with a relatively small $m$. 
The reason for this being true, and thus why the field of applicability of tensor network is so vast especially in one dimensional systems, stems from profound reasons related to area laws and the amount of quantum 
correlations that is possible to build with physical Hamiltonians~\cite{Eisert2010}. However, even if this was not true, a constructive approach is to assume that the state has a MPS description at given fixed $m$, 
and perform numerical simulations whose convergence can be checked as a function of increasing bond dimension $m$. Having introduced this approach -- choose a TN structure with fixed bond dimensions 
and use it as an ansatz to describe the many-body state -- it is easy to generalize it to different tensor structures. Indeed, according to the system to be described, different choices are possible, 
as sketched in Fig.~\ref{TNansatz} by means of a commonly used graphical representation of TN~\cite{Schollwock2011}. The generalization of MPS to two-dimensional system is known as Projected Entangled Pair States (PEPS)~\cite{Lubasch2014}; 
while mixed states can be described either by Matrix Product Operators (MPO) which also efficiently represent a large class of operators (many-body Hamiltonians or time evolution operators)~\cite{Verstraete2004},
 or by Locally Purified Tensor Networks (LPTN) which despite being slightly more complex than MPOs have the nice property of being positive defined by construction~\cite{Kliesch2014b,Cuevas2013,Werner2014}. Finally, one can also build hierarchical structures 
 which are intimately related to the real space renormalization group procedure,  such as Tree Tensor Networks (TTN)~\cite{Tagliacozzo2009,Evenbly2012,Gerster2014b} and Multi-scale Entanglement Renormalization Ansatz (MERA)~\cite{Vidal2007}.
Once the tensor network ansatz of the state is chosen, a whole family of algorithm, polynomially in the system size $N$, can be developed to 
study the properties of system under study within this approximation: one has access to the ground state properties and 
the time evolution determined by the time-dependent Schr\"odinger equation or by a Lindblad master equation in case of open systems~\cite{Schollwock2011,Daley2014,Bonnes2014,Werner2014}. 
Again, we stress that even though the tensor structure is fixed, the tensor entries and their auxiliary dimensions are free parameters: TN algorithms are build to optimize them to obtain the best possible 
approximation of the state of interest with the given (polynomial) resources. 
In some cases, especially for phenomena out of equilibrium, it is not guaranteed that an efficient tensor network description of the state exists; 
in others this might also be not possible. However, even in those worst cases, TN methods result in the best possible description compatible with the available resources, certified by an estimate of the errors which determine the level of confidence of the numerical simulations results.

\subsection{Gauge symmetries}

The TN presented in the last section could be used directly to study LGT as we will report later on, however it would be preferable in term of numerical efficiency and stability to exploit 
the symmetries present in the LGT. One possible way of doing that is to exploit the quantum link formulation of lattice gauge theories introduced in Sec. 2. 
In particular, we exploit the Schwinger representation introduced in 
Eq.\eqref{Sbosons} as it allows to satisfy the gauge symmetry on all lattice sites in a straightforward way. We specify what follows for one-dimensional Abelian LGT, however it can be generalized to 
the cases of non-Abelian and/or higher dimensional cases~\cite{Silvi2014}. Other approaches to simulate LGT via TN are possible, and we will recall some 
results obtained with alternative formulations in the next section, the interested reader is refereed to the existing exhaustive 
literature on the subject~\cite{Tagliacozzo2014,Banuls,Zohar2015,Notarnicola2015a,Osborne,Haegeman2015,Zohar2015a}. 

\begin{figure}[t]
\begin{center}
\begin{minipage}{160mm}
\resizebox*{16cm}{!}{\includegraphics{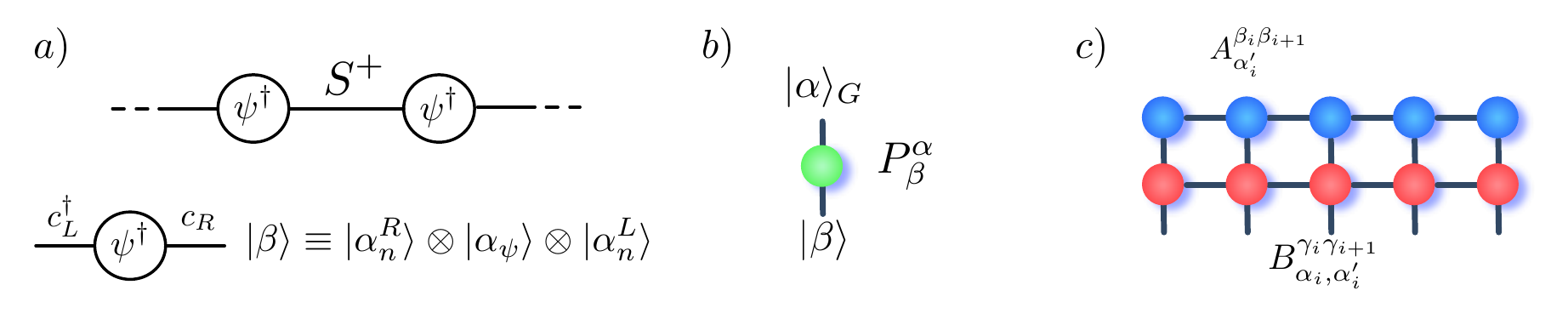}}\hspace{6pt}
\caption{Gauge invariant tensor networks. a) The quantum link representation of LGT defines the logical site local bases  $ |\alpha_n^R \rangle  | \alpha_\psi \rangle |\alpha_n^L \rangle$, composed by the matter degree of freedom on the lattice vertex 
and one rishon degree of freedom for each incoming link. b) The projector $P_{\alpha}^\beta = \hat P_{\alpha_n^R, \alpha_\psi, \alpha_n^L}^\alpha$, defined by the quantum link representation of the LGT, imposes the gauge symmetry defined by Eq.~\eqref{gcond1}. c) The gauge invariant many-body wave function can be expressed as a 
MPS (blue tensors) containing the variational parameters and a MPO (red tensors) with constant entries, which constraints the number of rishons per link 
$N_r$. 
} \label{LGTN}
\end{minipage}
\end{center}
\end{figure}

The first step to define a gauge invariant tensor network in the quantum link formulation is to recognize that the many-body Hilbert space written in the Schwinger representation can be build by tensor product of three local Hilbert spaces for each vertex of the lattice: one for the matter field 
$|\alpha_\psi \rangle = (\psi^\dagger)^\alpha |0\rangle$ and one for each rishon degree of freedom connected to it (two for 1d LGT), $ |\alpha_n^{L,R}\rangle =  (c^\dagger)^\alpha |0 \rangle $. Given that the 
the number of rishons per link is determined by the spin representation of the gauge fields, $N_r = 2 S +1$, the maximal occupation of the rishon degree of freedom is $N_r$. For example, for spin representation 
$S=1/2$ and spinless matter, the dimension of the Hilbert space of the ``logical site" is defined as 
$ | \ell \rangle = |\alpha_n^R \rangle \otimes | \alpha_\psi \rangle \otimes |\alpha_n^L \rangle$, is $d= 2 \times 2 \times2=8$ and $\ell = 1, \dots 8$.  
However, one can now impose the gauge constraint expressed as 
\begin{equation}
G_i - n_i - \frac{(-1)^i +1}{2} =\sum_{(i,j)\in -} (n_R-n_L )_{(i,j)} = (2 n_R - 2 N_r  + 2n_L)_{i}.
\label{gcond1}
\end{equation}
Given that all gauge generators $G_i$ have disjoint support 
(each of them only acts on a matter field on the vertex and on the rishon on the left and on the right of it) 
we can write the many-body Hamiltonian in a gauge invariant base. Indeed, as depicted in Fig.~\ref{LGTN}, projecting the local Hilbert space into the gauge invariant subspace, we arrive at 
\begin{equation}
| \alpha \rangle_G = \sum_{\alpha_n^R, \alpha_\psi, \alpha_n^L}  \hat P_{\alpha_n^R, \alpha_\psi, \alpha_n^L}^\alpha |\alpha_n^R \rangle  | \alpha_\psi \rangle |\alpha_n^L \rangle,
\end{equation}
where the number of gauge invariant states is $d'<d$, $\hat P$ is the corresponding projector over the subspace spanned defined by the gauge constraint $G_i |\alpha_i\rangle =0$, and for easy of notation we omitted the lattice site index $i$. 
Our previous example (spin representation $S=1/2$ and spinless matter) results in $d=3$. A general wave function beloging to the Hilbert space defined by the tensor product of $N$ gauge invariant 
Hilbert spaces can now be written easily as $|\psi \rangle = \sum_{\vec \alpha} \psi_{\vec \alpha} \prod_{i=1}^N  | \alpha_i \rangle_G$. As we have shown for the general case, the rank-N tensor $\psi_{\vec \alpha}$ 
can be replaced by a MPS. However, one can easily check that there are configurations in such MPS that violate the constraint of $N_r$ rishons per link. This constraint is an additional gauge symmetry introduced 
by the Schwinger representation that shall be satisfied. Notice that this additional gauge symmetry is Abelian independently from the nature of the original one, thus both for Abelian and non-Abelian LGT 
the constraint on the number of rishons per link can be imposed via local projectors acting on each link, diagonal in the local basis and thus commuting. The details of this construction are pretty technical, 
and while the interested reader can find them in~\cite{Silvi2014}, here we report only that it is possible to write this projector as a MPO. 
In conclusion, the resulting tensor structure --that defines the class of gauge symmetric tensor network in the quantum link representation--
takes the form of a MPS defined over the gauge invariant bases containing the variational parameters $A_{\alpha_i'}^{\beta_i\beta_{i+1}}$, and a MPO layer which imposes the constraint on the number of rishons per link, whose elements $B_{\alpha_i,\alpha_i'}^{\gamma_i\gamma_{i+1}}$ are defined by the gauge symmetry of the theory (see Fig.\ref{LGTN}): 
\begin{equation}
|\psi \rangle_\textrm{GTN} = \sum   A_{\alpha_1'}^{\beta_1}  A_{\alpha_2'}^{\beta_1\beta_2}    \dots A_{\alpha_N'}^{\beta_{N-1}}  
B_{\alpha_1,\alpha_1'}^{\gamma_1}  B_{\alpha_2,\alpha_2'}^{\gamma_1\gamma_2}  \dots B_{\alpha_N,\alpha_N'}^{\gamma_{N-1}}  
| \alpha_1 \alpha_2 \dots \alpha_N \rangle_G. 
\label{gTN}
\end{equation}
Starting from this assumption one can write efficient algorithms to compute ground states and time evolution properties of the LGT, generalizing the standard ones. 
In particular, it can be shown that exploiting the properties of the MPO in Eq.\eqref{gTN}, a speed up of the order of $N_r^2$ can be achieved~\cite{Silvi2014}. 

\subsection{Tensor network simulations of lattice gauge theories}

In this section we first report some results we obtained via the gauge invariant tensor method approach presented above, and conclude with a  brief excursion on results obtained via tensor network simulations of LGT via alternative approaches. 

The first application of gauge invariant TN has been the study of the phase diagram of the simplest non trivial lattice gauge theory with matter fields in the quantum link representation, the $U(1)$ gauge theory or QED in 1+1d presented in the previous sections~\cite{Rico2014}. There, we showed that the system undergoes to a quantum phase transition from a charge and parity ordered phase with nonzero electric flux to a disordered one with a net zero electric flux configuration and that this quantum phase transition is described by the Ising universality class. These findings have been obtained using both the spin one-half and spin one representation of the gauge field. 
We showed how it is possible to study not only local order parameters, quantities that can be extracted routinely 
via Monte Carlo calculations, but also one has access to the Von Neumann entropy of subsystems, 
which cannot be directly accessed via standard lattice gauge calculations. Our analysis has not been limited to equilibrium 
properties of the theory, but given the possibility of performing real-time evolution granted by the TN approaches, we also study the time evolution of out of equilibrium scenarios as the real-time dynamics of string breaking and of scattering~\cite{Pichler2015}.
We showed that both these phenomena are intimately related with entanglement generation: the Schwinger mechanism 
is tightly linked to entanglement spreading and the particle scattering generates entanglement among them. Finally, 
we determined a dynamical state diagram for string breaking, and we quantitatively evaluate the time-scales for mass production.

Alongside the aforementioned results, thriving research activities of different groups explored alternative approaches to the study of lattice gauge theory by means of TN methods. An alternative formulation of gauge invariant TN has been proposed and exploited to study ladder geometries of a $\mathbb{Z}_2$ LGT~\cite{Tagliacozzo2014}. Moreover, the relation between entanglement renormalization and gauge symmetries has been explored~\cite{Tagliacozzo2011}. 

The study of lattice gauge theory has been addressed successfully also by standard TN methods, without explicitly exploiting the gauge symmetries 
in the tensor formulation. Indeed, one can always check a posteriori the gauge invariance of the computed states: in a such a way,  
a $\mathbb{Z}_2$ LGT in a ladder and 2D geometries has been studied~\cite{Sugihara2005,Dusuel2011} and more recently insightful studies of the dynamics 
of string breaking in a $SU(2)$ LGT and a study of the continuum limit of the Schwinger model has been performed~\cite{Kuhn2014,Kuhn2015}. 
Exploiting the mapping of the Schwinger model to a long-range interacting spin system, one can also apply TN methods to perform detailed studies in one dimension, where many aspects of the Schwinger model in its Wilson formulation have been recently elucidated~\cite{Byrnes2002,Banuls,Buyens2014}, including the low-lying spectrum~\cite{Banuls:2013qf} and the chiral condensate fraction~\cite{Saito:2015kq,Banuls2015}.

Finally, alternative proposals to introduce LGT representations with finite local dimensions that naturally fits into TN and quantum simulations descriptions 
has been introduced by different authors together with studies on the theoretical properties of gauge symmetries and tensor networks~\cite{Zohar2015,Notarnicola2015a,Osborne,Haegeman2015,Zohar:2015yq,Zohar:2015bf}.

\section{Quantum simulation of lattice gauge theories}
\label{QsymLGT}

The realization of LGTs in synthetic quantum systems presents a very specific challenge with respect to other statistical mechanics models such as spin and Hubbard-type Hamiltonians, that is, the presence of a series of constraints described by the Gauss's laws. In the following, we first review few strategies to realize such constrained dynamics, and then illustrate some examples presented in literature, focusing mostly on the conceptually simpler Abelian theories. We will present most of the results at a qualitative level, referring the reader to the original literature, and to two specialized reviews~\cite{Wiese:2013kk,Zohar2015a} which extensively cover many of the discussed topics.

\subsection{How to engineer gauge symmetries: some strategies}

Usually, quantum simulator platforms describe Hamiltonian dynamics on a Hilbert space where only global symmetries are present. A prototypical example is fermionic atoms hopping on an optical lattice, described by the Hamiltonian in Eq.~\eqref{FF_ham}: here, the only symmetry is a global U(1), related to particle number conservation (in addition to discrete lattice symmetries). At first sight, one could think of directly engineering the more complicated interactions of the kind of Eq.\eqref{Htt} present in gauge theories: however, frequently other terms which spoil gauge invariance would be present, thus making the {\it brute force} approach not always viable. 

As said previously, the key point in devising implementations for LGTs is to act at the Hilbert space level: while quantum systems  generically act on the full combined Hilbert space $\mathcal{H}=\mathcal{H}_P\otimes\mathcal{H}_{U}$, it is possible to identify interaction engineering schemes that allow to integrate out the dynamics in the unphysical subspace (or to neglect it completely), thus generating an effective theory acting solely on the physical part of the Hilbert space. 

\subsubsection{Energy punishment}
 
A standard way to engineer constraints, widely used in the field of frustrated quantum magnets~\cite{Lacroix2010}, is to penalize all gauge variant states in energy, such that the low-energy physics takes place only on the gauge invariant Hilbert space. Given a set of generators $\{G_x\}$, such that $[G_x,G_y] = 0$, one starts with Hamiltonians of the form:
\begin{equation}
H_{Ep} = V H_0+\lambda H_1, \quad H_0=\sum_{x} G_x^2 , \quad V>0
\label{enpush}
\end{equation}
so that, in the limit $V\gg \lambda$, all states which do not satisfy Gauss's law have energy $E\geq V$. The low-energy dynamics is then driven by $H_1$, which can either be gauge invariant, or gauge variant. In the latter case, the gauge invariant Hamiltonian is then generated in perturbation theory. In the case of quantum Ice Hamiltonians one has:
\begin{equation}
H_0 = \sum_{+}(\sum_{j\in+}S^z_j)^2
\end{equation}
while $H_1$ is typically of the form:
\begin{equation}
H_1 = h\sum_{j}S^x_j + J \sum_{<i,j>}(S^+_iS^-_j+\textrm{h.c.})
\end{equation}
so that, using either fourth-order perturbation theory in $h/V$ or second-order in $J/V$, one realizes the desired effective Hamiltonian $H_{QI}$ (up to corrections of order $\mathcal{O}[(h/V)^8]$ and $\mathcal{O}[(J/V)^4]$, respectively). 

The energy penalty strategy can be easily applied to Abelian gauge theories; in fact, this is typically employed in condensed matter studies to identify microscopic Hamiltonians described at low-energies by quantum dimer and quantum loop models~\cite{Lacroix2010,wenbook,fradkinbook}. However, in the non-Abelian case, the situation is considerably more complicated, as the generators do not commute, and thus one has to identify a proper microscopic symmetry in order to avoid fine-tuning problems.

\subsubsection{Quantum Zeno dynamics}

While usually very well isolated from any environmental noise, synthetic quantum systems offer the possibility of using tailored dissipation to engineer a dissipative quantum dynamics~\cite{Diehl:2008xy,Verstraete:2008sf}. The main idea is that one can introduce controlled noise in the system, either via external fields (e.g., a noise laser field) or by exploiting intrinsic dissipation channels (such as three-body losses~\cite{Kantian:2009rw,Daley:2009qr}). This strategy has already been applied in different contexts, ranging from dissipative state preparation~\cite{Diehl:2008xy,Verstraete:2008sf,Syassen:2008it,Garcia-Ripoll:2009ek} to entanglement generation via dissipation~\cite{Muschik:2012fv}. 

A particularly useful engineered dissipation in the context of constraint engineering is the so called quantum Zeno dynamics~\cite{Facchi:2000bs,Facchi:2002ij}. This mechanism is directly inspired by the quantum Zeno effect, which states that a system continuously observed does not evolve as a function of time. In the case where part of the Hamiltonian dynamics commutes with the external perturbation introduced by the measurement, a non-trivial Quantum Zeno dynamics can take place within a subspace of the total Hilbert space~\cite{Facchi:2002ij}. In Ref.~\cite{Stannigel:2014bf}, it was shown how this effect can be used to engineer the constraints corresponding to the Gauss's law in both Abelian and non-Abelian cases. The starting point is to consider a time-dependent Hamiltonian:
\begin{equation}
\label{eq:Hclassicalnoise}
H_{\rm noise}(t)=H_{1}+H_{2}+\sqrt{2\kappa}\sum_{x,a}\xi_{x}^{a}(t)G_{x}^{a}\,, 
\end{equation}
where the $H_1 (H_2)$ are time-independent gauge invariant (variant) contributions, while the last term describes classical noise terms $\xi^a_x(t)$ individually coupled to the generators of the gauge symmetry $G_x^a$. Averaging over the noise sources and assuming white noise correlations, one gets a description in terms of a master equation~\cite{gardiner_book}:
\begin{align}
\label{eq:ME}
\dot{\rho}=-iH_{\mathrm{eff}}\rho+i\rho H_{\mathrm{eff}}^{\dag}+2\kappa
\sum_{x,a}G_{x}^{a}\rho G_{x}^{a}\,,
\end{align}
where  $\rho$ is the density matrix of the system and with non-Hermitian Hamiltonian $H_{\mathrm{eff}}$ given by 
\[
H_{\mathrm{eff}}=H_{0}+H_{1}-i\kappa\sum_{x,a}(G_{x}^{a})^{2}\,.
\]
The effective Hamiltonian contains a damping term which is quadratic in the symmetry generators, very much akin to the energy punishment introduced in Eq.~\eqref{enpush}. An analysis in second order perturbation theory in $\kappa/J$ (where $J$ is the leading energy scale in $H_2$) leads to an effective dynamics constrained to the gauge invariant Hilbert space, and described by the effective Hamiltonian:
\begin{equation}
\tilde{H}_{\mathrm{eff}}\approx\mathcal{P}(H_{0}+H_{1})\mathcal{P}-i\,\mathcal{P}%
H_{1}\mathcal{Q}\frac{1}{\kappa\sum_{x,a}(G_{x}^{a})^{2}}\mathcal{Q}%
H_{1}\mathcal{P}\, \label{eq:Heff2}
\end{equation}
with $\mathcal{P}$ and $\mathcal{Q}$ being projector operators on $\mathcal{H}_P$ and $\mathcal{H}_{U}$, respectively. This strategy can be applied to both abelian and non-Abelian gauge groups, as long as the proper noise source can be identified. Notice that contrary to the energy penalty strategy, leakage out of $\mathcal{H}_{P}$ will take place as a function of time, albeit at a very small rate $\propto J^2/\kappa$.

\subsubsection{Exploiting microscopic symmetries}

Another strategy to engineer local symmetries in synthetic systems is to exploit microscopic symmetries of (atomic or molecular) matter, and to combine those with spatially arranged potentials in such a way that the emerging dynamics is gauge invariant up to very high energy scales. This strategy has been first considered in Refs.~\cite{Banerjee2013,Zohar:2013qf,Zohar:2013kb}, where scattering symmetries (either angular momentum conservation~\cite{Zohar:2013qf,Zohar:2013kb}, or SU(N) symmetries~\cite{Banerjee2013} of alkaline-earth-like atoms within their ground state manifold) were exploited. The basic idea is that one considers minimal building blocks, where a scattering symmetry is thus global, and carefully glues them altogether, so that the dynamics of the combined system still preserves the full set of local conservation laws of the building blocks. As this strategy very much builds upon the specific implementation scheme, we will illustrate it in the following section using a concrete cold atom example from Ref.~~\cite{Zohar:2013qf,Kasper:2015kc}.

\subsubsection{Digital quantum simulators}

In addition to the analog quantum simulation strategy outlined above, it is possible to devise specific protocols to engineer gauge theories on quantum computers, or digital quantum simulators~\cite{lloyd1996universal}. The main idea at the heart of digital quantum simulation is to decompose the (local) Hamiltonian evolution operator $e^{-iHt}$ of interest into small contributions involving a limited number of fields using the Trotter formula:
\begin{equation}
e^{-itH} \simeq \left(\prod_{\alpha=1}^{M}e^{-iH_\alpha t/n}\right)^n, \quad H = \sum_{\alpha=1}^M H_\alpha
\end{equation}
and then to interpret the small steps as quantum gates acting on a register of qubits. In this respect, the differences between the realization of gauge theories and conventional spin models are milder than in the analog approach outlined above: there is no need to directly impose constraints, as the Hamiltonian is gauge invariant by definition. The only additional difficulty is that multi-spin interactions have to be decomposed into a sequence of gates in an efficient way. Several proposals have been outlined in literature, including both Abelian~\cite{Tagliacozzo:2012kq,Weimer:2010xw} and non-Abelian theories~\cite{Tagliacozzo2013,Egusquiza2015}, and can be applied in an almost platform independent way.

\subsection{Cold atoms in optical lattices}

Cold atoms trapped in optical lattices are one of the leading technologies for the realization of synthetic quantum systems combining accurate interaction engineering with control at the single quantum level~\cite{Bloch2008,Bloch2012}. Starting from the observation of the superfluid to Mott insulator transition in a gas of $^{87}$Rb atoms in an optical lattice~\cite{greiner2002quantum}, proposed in Ref.~\cite{jaksch1998cold}, many paradigmatic system Hamiltonians have been realized in these systems, including SU(2)~\cite{jordens2008mott,schneider2008metallic} and SU(N) fermionic Hubbard models~\cite{Fallani14,Taie:2012ys,Hofrichter:2015fv}, long-range interactions~\cite{Paz:2013dn,Yan:2013bh,Schaus:2015it}, and static gauge fields~\cite{Atala:2014th,Mancini25092015,Miyake:2013dz,Struck:2013yf,Stuhl:2015rw}, just to name a few. 

The key properties of cold atom settings is the ample degree of flexibility and tunability~\cite{Jaksch2005}: various lattice geometries can be realized; both Fermi and Bose gases can be loaded into the lattices themselves~\cite{Bloch2008,Bloch2012}; the inter particle interactions can be tailored by using external fields either using the so-called Feshbach resonances~\cite{Chin:2010fe}, or by exploiting naturally occurring long-range interactions between magnetic or electric dipoles~\cite{Baranov:2012wo,Lahaye:2009by}, or via Rydberg states~\cite{Saffman:2010bq}. For a review on the current status in the fields, see Ref.~\cite{Bloch2012}. 

\subsubsection{Gauge theories coupled to matter fields}

Owing to their ample flexibility, cold atoms are a premier candidate for the physical realization of dynamical gauge fields coupled to fermionic matter. All of the strategies discussed above have been employed to devise implementation schemes for both Abelian~\cite{Kapit:2010kh,Banerjee2012,Zohar:2013om,Bazavov:2015zh,Notarnicola2015a,Kasper:2015kc} and non-Abelian~~\cite{Banerjee2013,Zohar:2013kb,Stannigel:2014bf} lattice gauge theories in the presence of matter fields. Here, we briefly review two examples for the U(1) case.

{\it U(1) lattice gauge theories coupled to matter via energy penalty.--}
In Ref.~\cite{Banerjee2012}, an implementation of the quantum link model in Eq.~\eqref{QLM_U1} at strong coupling (e.g., with $V=J=0$) was presented. The starting point is the realization of the gauge fields on the bonds. This is done by considering bosonic double wells hosting bosons of two species $b^\sigma_{L/R}$, $\sigma=1,2$, in a staggered pattern. In each well, the total number of bosons $N=n^\sigma_R+n^\sigma_L$, is fixed, thus realizing a effective spin $S= N/2$ degree of freedom:
\begin{equation}
S^z_{x,x+1} = (b^{1\dagger}_{x,x+1; L} b^{1}_{x,x+1; L} - b^{1\dagger}_{x,x+1; R} b^{1}_{x,x+1; R})/2,\quad S^+_{x,x+1} = b^{1\dagger}_{x,x+1; L} b^{1}_{x,x+1; R}
\end{equation}
on the even bonds, and the same replacing $1\rightarrow 2$ for the bosonic index $\sigma$ on the odd bonds. The fields $b^\sigma_{L/R}$ are thus interpreted as Schwinger bosons, whose microscopic dynamics is described by a tunneling Hamiltonian between the different sites of each double well, that is $H_B = -t_B\sum_{x}S^x_x$.
The dynamics of an additional  spinless fermion species $\psi_x$ is described by the Hamiltonian:
\begin{equation}
H_F = -t_F\sum_{x}(\psi^\dagger_x\psi_{x+1}) +\sum_{x}(-1)^xn_x
\end{equation}
with $n_x=\psi^\dagger_x\psi_x$. In order to ensure gauge invariance, an additional term $H_U = U\sum_{x}G_x^2$ is introduced, with the generators $G_x$ defined as:
\begin{equation}
 G_x = n_x^F + n^1_{x, x+1, R} + n^2_{x+1, x+2, L} - 2S +
\frac{1}{2}\left[(-1)^x - 1 \right].
\end{equation}
for the even-odd bonds. This can be generated using a combination of local interactions and superlattice structures, as described in detail in Ref.~\cite{Banerjee2012}. The emergent dynamics can then be obtained using degenerate second order perturbation theory, where the QLM in Eq.~\eqref{QLM_U1} is recovered, with couplings $t\propto(t_Ft_B)/U$. In Ref.~\cite{Banerjee2012}, it has been shown how this minimal realization of a lattice gauge theory already host paradigmatic physical phenomena, in direct connection to the Schwinger model. In particular, in these cold atom realizations, one could study the real-time dynamics of string breaking, which reveals the interesting dynamics of confining theories in the presence of dynamical matter. Recently, extension of this scheme in the context of $\mathbb{Z}_N$ LGTs coupled to matter fields have also been proposed, which could potentially connect to the U(1) case in the large $N$ limit~\cite{Notarnicola2015a}. 

Moreover, simple scattering experiments along the lines of Ref.~\cite{Pichler2015} could also be investigated: as a first step, one could adiabatically prepare the system into some excited states, e.g., a pair of mesons separated by a vacuum region. Subsequently, ramping down the optical lattice for the fermions would quench the Hamiltonian from $t_{i} = 0$ to $t_{f}>0$, giving rise to scattering between the mesons. Finally, it is important to note that the relevant timescales for these experiments range in the 10 to 100 Hz range, that is, are the same as the ones required to investigate the physics of quantum magnetism via super-exchange interactions~\cite{Bloch2008,Bloch2012}.

\begin{figure}[t]
\begin{center}
\begin{minipage}{80mm}
\resizebox*{8cm}{!}{\includegraphics{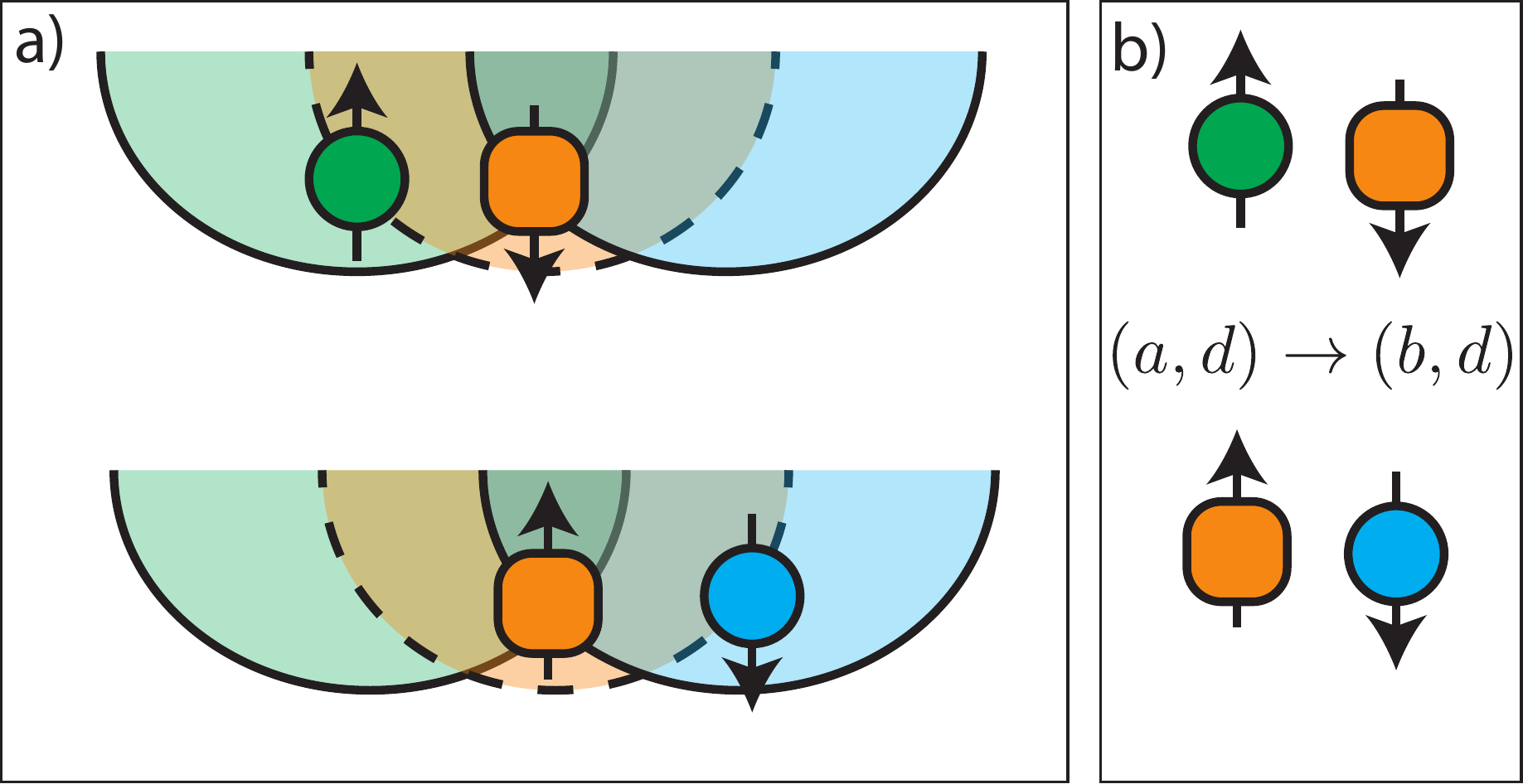}}\hspace{6pt}
\caption{Schematics of spin-changing collisions leading to effective gauge matter interactions. Panel a);The fermions are represented by circles, and are subject to a state-dependent lattice; the bosons are represented by boxes, and both spin states are trapped at the same site residing on the bonds between each pair of fermions. The scattering process described in b) drives a fermion tunneling while at the same time flips the spin of the boson at the link.
} \label{Spinchanging}
\end{minipage}
\end{center}
\end{figure}

{\it U(1) lattice gauge theories coupled to matter via spin-changing collisions.--}
An alternative way to generate QLM dynamics is to employ spin-changing collisions~\cite{Zohar:2013qf,Kasper:2015kc}. The basic idea is illustrated in Fig.~\ref{Spinchanging}. Bosons in two internal states $a,b$ with different spin $m_F, m_F+1$ are confined in a deep lattice, and serve as effective link variables, where the spin degree of freedom is again expressed as product of Schwinger bosons, $S^+ = a^\dagger b$. An additional fermionic species with two internal states $c, d$ is then introduced, and is subject to a state dependent lattice which traps $c$ (d) on even (odd) sites. The lattice is supposed to be sufficiently deep that tunneling of the fermions (without any spin flip) is suppressed over the timescales of the experiment.

Since particles cannot tunnel due to an energetic constraint, the only residual interaction left is induced via spin-changing collisions. In particular, tuning the magnetic field such that the Zeeman splittings between $a,b$ and $c,d$ are equal, one gets a resonant spin-changing interaction of the form:
\begin{equation}
(a, d) \rightarrow (b, d).
\end{equation}
Since the Wannier functions of both bosons and fermions are strongly localized at each lattice site, using the identification $c_i =\psi_{2i}, d_{i} = \psi_{2i+1}$, naturally leads to terms of the form $\psi^\dagger_{2i}S^+_{2i, 2i+1}\psi_{2i+1}$, exactly the ones required for the realization of the U(1) QLM. Similar methods using combination of spin changing collisions with either energy punishment or Zeno dynamics have also been discussed in the context of non-Abelian lattice gauge theories, in particular in Refs.~\cite{Zohar:2013kb,Stannigel:2014bf}, and for the investigation of QLMs approaching the Wilson formulation~\cite{Kasper:2015kc}.

\subsubsection{Extensions to non-Abelian theories and approaching the continuum limit}

As we have seen in Section~\ref{IntroLGT}, the formalism introduced for the Abelian case can be generalized to non-Abelian symmetries. Implementation schemes for the realization of both SU(2)~\cite{Zohar:2013kb,Tagliacozzo2013,Egusquiza2015} and U(N)/SU(N)~\cite{Banerjee2013,Stannigel:2014bf} LGTs in cold atom systems have been proposed. As the realization of non-Abelian symmetries is conceptually more challenging than the Abelian counterpart described above, we refer the reader to the original references for a full explanation of the models and their cold atom realizations. Another relevant example of the flexibility of cold atom systems is the possibility of engineering lattice models which can systematically approach the continuum limit in the context of the so-called D-theories. In particular, in Refs.~\cite{Laflamme:2015oe,Laflamme:2015sy}, a scheme was proposed where $\mathbb{C}P(N-1)$ models, a paradigmatic example of (1+1)d field theory which originates from a U(1) gauge symmetry, can emerge as an effective low-energy description of certain SU(N) quantum spin ladders. Despite their simplicity, these models share fundamental features with QCD, such as asymptotic freedom and confinement~\cite{ALV}, which could be probed in cold atoms laboratories using already developing measurement tools. 

\subsubsection{Quenched gauge theories}

While applications in the context of high-energy physics usually require the presence of dynamical matter, many quenched gauge theories emerge in the description of frustrated quantum magnets, either as effective low-energy descriptions, or as paradigmatic model Hamiltonians. In this context, possible implementations of spin Ice Hamiltonians  have been proposed using different platforms, both analog and digital ones. In Ref.~\cite{Buchler2005}, an implementation of the ring exchange term $H_\square$ has been proposed using bosonic atoms in an optical lattice. There, the main ingredient is to use molecular state on the dual lattice to generate the ring-exchange term using perturbation theory, and to suppress single particle tunneling via quantum-interference effects. The ring exchange term also has natural realization using digital implementation schemes, as extensively discussed in Refs.~\cite{Tagliacozzo:2012kq,Weimer:2010xw}, and using Bose-Einstein condensates in state-dependent lattices~\cite{Zohar:2012kl}.

A generic platform for quantum dimer models on different geometries is provided by systems with long-range (or just non-local) interactions, using the energy punishment strategy. In Ref.~\cite{Tewari:2006ss}, a gas of polar molecules confined on Kagome and pyrochlore lattices was considered. Using a combination of on-site and off-site repulsion, it was shown how an emergent quantum dimer model can be built in the dual lattice of such geometries, that is, on the honeycomb and diamond lattice. An even more attractive route is provided by Rydberg atoms. In these systems, the interactions can be specifically tailored in space using different techniques, such as Rydberg dressing, or using light-fields to properly engineer the Rydberg manifold in order to promote certain interactions channels as dominants~\cite{Glaetzle2014,Bijnen:2015bx,Glaetzle2015}. A particularly simple example were such interactions are needed, which cannot be achieved using long-range interactions only, is quantum Ice on the square lattice~\cite{Glaetzle2014}. In this case, both anisotropic and position dependent interactions are needed: both features can be engineered  using Rydberg dressing to p-states. Within the same toolbox, a variety of other quantum dimer models, spanning various lattice geometries such as Kagome and 4-8 lattice, can also be realized.

\subsection{Trapped ions}

The impressive technological advances in trapped ion systems experienced in the last decades, made them a paradigm for both quantum information processing and, more recently, quantum simulation~\cite{blatt2012quantum}. In the context of lattice gauge theories, some proposals have been recently put forward using both ion chains and 2D Coulomb crystals.

Ion chains are ideal settings to investigate the physics of (1+1)d field theories (see, e.g., the pioneering work on the Dirac equation~\cite{gerritsma2010quantum}). Usually, the dynamics in these systems is described by the interplay between internal level of the ions, the phonon fields of the ion crystal, and the light coupling the ion motion to their internal structure. This results in Hamiltonians of the spin-boson- or spin-type~\cite{Porras:2004km}, with interaction tunability typically provided by the use of external light fields~\cite{schneider2011many}. In Ref.~\cite{Hauke2013}, it is shown how the dynamics of the Schwinger model in the quantum link formulation can be realized using such interactions, using as matter and gauge field degrees of freedom the internal states of the ions. In particular, thanks to a mapping to monotonic interactions which allow to design a simple energy penalty term for gauge invariant states, the full phase diagram of the Schwinger model can be probed experimentally, in addition to its time-dependent dynamics. 

Another typical platform for trapped ions systems are two-dimensional Coulomb crystals: due to the strong inter-ion repulsion, ions strongly confined to a plane geometry form triangular shaped crystals, on the top of which the desired dynamics can be imprinted using the same light-matter interactions exploited for 1D chains. The triangular geometry provides then a starting point to engineer various models of quantum frustration with close ties to emergent gauge theories~\cite{Lacroix2010}. An example is discussed in Ref.~\cite{Nath:2015to}, where a sub lattice of ions was dynamically hidden. This allows one to effectively create a Kagome lattice, with the hidden ions located at the center of each hexagonal plaquette. These decoupled ions are then used to tailor the lattice band structure to support `localized' excitations at the center of each hexagonal plaquette, either using an optical pinning laser, or via Rydberg dressing. The emergent interactions involve ions belonging to the same hexagonal plaquette, thus realizing the so-called Balents-Fisher-Girvin model~\cite{Balents:2002yi}. This model, which is the Kagome lattice version of a U(1) QLM, is known to support a $\mathbb{Z}_2$ spin liquid phase, separated from a superfluid phase by a deconfined quantum critical point~\cite{Isakov:2012wl}.

\subsection{Superconducting circuits}

Arrays of superconducting qubits have been shown to be promising candidates for the realization of strongly correlated phases of matter, ranging from quantum spin systems to a variety of Hubbard models~\cite{Houck:2012,Viehmann2013a,Mezzacapo:2014bs,Naether:2014dz}. Owing to impressive improvements in fabrication technology, these solid state devices have ample potential in terms of both scalability and flexibility design of their basic components. As in the ion case, these systems feature mostly bosonic and spin degrees of freedom, thus making them ideal playgrounds for both low-dimensional or pure gauge theories. In particular, implementation schemes for both Abelian and non-Abelian gauge theories have been put forward~\cite{Doucot:2004rm,Marcos2013,marcos2014two,Egusquiza2015}. In Refs.~\cite{Marcos2013,marcos2014two}, a toolkit for the realization of the U(1) QLM has been described. A particular interesting features of circuit QED architectures is that various geometries can be realized, which in the context of gauge theories opens up the possibility of studying string-breaking and string excitations with different boundary conditions, describing static charges at the edge.

In Ref.~\cite{Doucot:2004rm,Egusquiza2015}, it has been shown how non-Abelian quenched gauge theories can also be realized using superconducting qubit architectures. In Ref.~\cite{Doucot:2004rm}, models for discrete non-Abelian symmetries, such as $S_3$, are introduced in the context of Josephson junction arrays. The main motivation behind the implementation of those models is their potential for quantum computing applications, along similar lines discussed by Kitaev using Abelian LGTs~\cite{Kitaev2003}. Continuos non-Abelian symmetries, in particular SU(2), are discussed in Ref.~\cite{Egusquiza2015}, using instead a digital quantum simulation scheme. A triangular lattice model is investigated in detail, and in particular, the quality of gauge invariance as a function of the Trotter time-step is shown to improve very quickly when decreasing the time-step itself. 

\section{Conclusions and outlook}
\label{conclu}

Gauge theories represent one of the fundamental building block for our understanding of natural phenomena, ranging from particle physics to condensed matter systems. They provide an extremely elegant formulation of physical laws stemming from the concept of symmetry, which many times allows for very general prediction of their properties. However, from a computational perspective, they are an archetypical example of strongly correlated systems, which challenges the capabilities of classical computational schemes. In this review, we have discussed how concepts and ideas from quantum information provide a novel route for the investigation of gauge theories, complementary to widely-used approaches based on Monte Carlo methods. 

Tensor network methods, which have found widespread application in condensed matter theory, have already been applied to the study of low-dimensional lattice gauge theories. While their application to real-world (3+1)-d gauge theories is a long-term goal, which will require theoretical breakthroughs along the way, it has already been shown how these methods compare favorably with Monte Carlo simulations in static cases where both approaches are applicable. Moreover, the possibility of investigating quench dynamics, where Monte Carlo typically suffers from complex action problems invalidating importance sampling, opens the way to a full-fledged quantum treatment of low-dimensional gauge theories in real time. 

From the quantum simulation viewpoint, the realization of gauge theories in synthetic quantum matter is a rapidly developing paradigm, which is just taking its first steps. From the theoretical side, many questions require additional work, such as understanding the typical error sources in the quantum simulation of gauge theories (such as coupling to Higgs fields) and the development of simpler strategies to realize gauge invariant dynamics. From the experimental side, while dynamical gauge fields are harder to get at work in the laboratories with respect to their static counterparts, we are optimistic that the first steps can be taken with current technologies in all available platforms.  

Finally, let us emphasize that the approaches outlined here are by no means alternative to the already developed computational methods employed for the investigation of lattice gauge theories, and which have led to remarkable predictions in the field of quantum Chromodynamics. Tensor network methods work at their best in regimes which are qualitatively different from the typical regimes of applicability of MCs: computationally speaking, the methods are almost fully complementary. Quantum simulation requires a major theoretical input for validation and benchmarking, for which realistic simulations are a necessary tool. In this sense, the developments summarized in this review shall hopefully serve to establish a novel interfaces between rather distant fields, so that novel methods and technologies can be developed, and further physical insights can be gathered from both perspectives.

We acknowledge discussions with D. Banerjee, M.~Burrello, J.I.~Cirac, P.~Hauke, V. Kasper, T.~Pichler, E.~Rico, P.~Silvi, F.~Tschirsich, U.-J.~Wiese, and P.~Zoller, and support from EU projects SIQS, RYSQ, the ERC Synergy Grant UQUAM, DFG SFB/TRR21, and the SFB FoQuS (FWF Project No. F4016-N23). SM would like to thank IQOQI for the kind periodic hospitality during which this paper has been written and C. Montangero for providing a computer scientist feedback to the manuscript.

\bibliographystyle{tCPH}
\bibliography{librarySMv3}

\begin{thebibliography}{100}
\newcommand{\noopsort}[1]{}
\newcommand{\printfirst}[2]{#1}
\newcommand{\singleletter}[1]{#1}
\newcommand{\switchargs}[2]{#2#1}
\providecommand{\url}[1]{\normalfont{#1}}
\providecommand{\urlprefix}{Available at }

\bibitem{Szabo1996}
A. Szabo and N.S. Ostlund, \emph{{Modern Quantum Chemistry: Introduction to
  Advanced Electronic Structure Theory}}, 1996.

\bibitem{Nielsen2011}
M.a. Nielsen and I.L. Chuang, \emph{{Quantum Computation and Quantum
  Information: 10th Anniversary Edition}}, 2011.

\bibitem{Kurizki2015}
G. Kurizki, P. Bertet, Y. Kubo, K. M{\o}lmer, D. Petrosyan, P. Rabl, and J.
  Schmiedmayer, \emph{{Quantum technologies with hybrid systems}}, Proc. Natl.
  Acad. Sci.  (2015), p. 201419326,
  \urlprefix\url{http://www.pnas.org/lookup/doi/10.1073/pnas.1419326112}.

\bibitem{Cheng2009}
Y.C. Cheng and G.R. Fleming, \emph{{Dynamics of light harvesting in
  photosynthesis.}}, Annu. Rev. Phys. Chem. 60 (2009), pp. 241--62,
  \urlprefix\url{http://www.ncbi.nlm.nih.gov/pubmed/18999996}.

\bibitem{Scholes2011}
G.D. Scholes, G.R. Fleming, A. Olaya-Castro, and R. van  Grondelle,
  \emph{{Lessons from nature about solar light harvesting.}}, Nat. Chem. 3
  (2011), pp. 763--74,
  \urlprefix\url{http://www.ncbi.nlm.nih.gov/pubmed/21941248}.

\bibitem{Buchmuller2006}
W. Buchm{\"{u}}ller and C. L{\"{u}}deling, \emph{{Field Theory and Standard
  Model}}, arXiv  (2006), \urlprefix\url{http://arxiv.org/abs/hep-ph/0609174}.

\bibitem{Lee2006}
P.a. Lee, N. Nagaosa, and X.G. Wen, \emph{{Doping a Mott insulator: Physics of
  high-temperature superconductivity}}, Rev. Mod. Phys. 78 (2006), pp. 17--85,
  \urlprefix\url{http://link.aps.org/doi/10.1103/RevModPhys.78.17}.

\bibitem{Mann2011}
A. Mann, \emph{{High-temperature superconductivity at 25: Still in suspense.}},
  Nature 475 (2011), pp. 280--282.

\bibitem{top500}
\emph{{Top 500 list}}, \urlprefix\url{www.top500.org}.

\bibitem{Wilson1975}
K. Wilson, \emph{{The renormalization group: Critical phenomena and the Kondo
  problem}}, Rev. Mod. Phys. 47 (1975), pp. 773--840,
  \urlprefix\url{http://rmp.aps.org/abstract/RMP/v47/i4/p773{\_}1}.

\bibitem{Hatting2009}
C. H{\"{a}}tting, \emph{{Electronic Structure : Hartree-Fock and Correlation
  Methods}}, Vol.~42, 2009.

\bibitem{rubinstein2011}
R.Y. Rubinstein and D.P. Kroese, \emph{{Simulation and the Monte Carlo
  method}}, 2011.

\bibitem{white1992}
S.R. White, \emph{{Density matrix formulation for quantum renormalization
  groups}}, Phys. Rev. Lett. 69 (1992), pp. 2863--2866,
  \urlprefix\url{http://prl.aps.org/abstract/PRL/v69/i19/p2863{\_}1}.

\bibitem{Schollwock2011}
U. Schollw{\"{o}}ck, \emph{{The density-matrix renormalization group in the age
  of matrix product states}}, Ann. Phys. (N. Y). 326 (2011), p.~96,
  \urlprefix\url{http://www.sciencedirect.com/science/article/pii/S0003491610001752
  http://arxiv.org/abs/1008.3477
  http://linkinghub.elsevier.com/retrieve/pii/S0003491610001752}.

\bibitem{Troyer2005}
M. Troyer and U.J. Wiese, \emph{{Computational Complexity and Fundamental
  Limitations to Fermionic Quantum Monte Carlo Simulations}}, Phys. Rev. Lett.
  94 (2005), p. 170201,
  \urlprefix\url{http://link.aps.org/doi/10.1103/PhysRevLett.94.170201}.

\bibitem{Feynman1982}
R.P. Feynman, \emph{{Simulating physics with computers}}, Int. J. Theor. Phys.
  21 (1982), pp. 467--488,
  \urlprefix\url{http://link.springer.com/10.1007/BF02650179}.

\bibitem{Cirac2012}
J.I. Cirac and P. Zoller, \emph{{Goals and opportunities in quantum
  simulation}}, Nat. Phys. 8 (2012), pp. 264--266,
  \urlprefix\url{http://www.nature.com/doifinder/10.1038/nphys2275}.

\bibitem{Bloch2012}
I. Bloch, J. Dalibard, and S. Nascimb{\`{e}}ne, \emph{{Quantum simulations with
  ultracold quantum gases}}, Nat. Phys. 8 (2012), pp. 267--276,
  \urlprefix\url{http://www.nature.com/doifinder/10.1038/nphys2259}.

\bibitem{blatt2012quantum}
R. Blatt and C. Roos, \emph{Quantum simulations with trapped ions}, Nature
  Physics 8 (2012), pp. 277--284.

\bibitem{Preskill2012}
J. Preskill, \emph{{Quantum computing and the entanglement frontier}}  (2012),
  pp. 1--18, \urlprefix\url{http://arxiv.org/abs/1203.5813}.

\bibitem{Bloch2008}
I. Bloch, J. Dalibard, and W. Zwerger, \emph{{Many-body physics with ultracold
  gases}}, Rev. Mod. Phys. 80 (2008), pp. 885--964,
  \urlprefix\url{http://rmp.aps.org/abstract/RMP/v80/i3/p885{\_}1}.

\bibitem{Jaksch2005a}
D. Jaksch and P. Zoller, \emph{{The cold atom Hubbard toolbox}}, Ann. Phys. (N.
  Y). 315 (2005), pp. 52--79,
  \urlprefix\url{http://www.sciencedirect.com/science/article/pii/S0003491604001782}.

\bibitem{Lieb1993a}
E. Lieb, \emph{{The Hubbard Model: Some Rigorous Results and Open Problems}}
  1993 (1993), p.~20, \urlprefix\url{http://arxiv.org/abs/cond-mat/9311033}.

\bibitem{Buchler2005}
H.P. B{\"{u}}chler, M. Hermele, S.D. Huber, M.P.a. Fisher, and P. Zoller,
  \emph{{Atomic Quantum Simulator for Lattice Gauge Theories and Ring Exchange
  Models}}, Phys. Rev. Lett. 95 (2005), p. 40402,

\bibitem{Weimer2010}
H. Weimer, M. M{\"{u}}ller, I. Lesanovsky, P. Zoller, and H.P. B{\"{u}}chler,
  \emph{{A Rydberg quantum simulator}}, Nat. Phys. 6 (2010), pp. 1--7,
  \urlprefix\url{http://dx.doi.org/10.1038/nphys1614}.

\bibitem{Tagliacozzo2013}
L. Tagliacozzo, A. Celi, P. Orland, M.W. Mitchell, and M. Lewenstein,
  \emph{{Simulation of non-Abelian gauge theories with optical lattices}}, Nat.
  Commun. 4 (2013), pp. 1--8,
  \urlprefix\url{http://www.nature.com/doifinder/10.1038/ncomms3615}.

\bibitem{Glaetzle2014}
A.W. Glaetzle, M. Dalmonte, R. Nath, I. Rousochatzakis, R. Moessner, and P.
  Zoller, \emph{{Quantum Spin-Ice and Dimer Models with Rydberg Atoms}}, Phys.
  Rev. X 4 (2014), p. 041037,
  \urlprefix\url{http://link.aps.org/doi/10.1103/PhysRevX.4.041037}.

\bibitem{Schauss2015a}
P. Schau{\ss}, J. Zeiher, T. Fukuhara, S. Hild, M. Cheneau, T. Macr{\`{\i}}, T.
  Pohl, I. Bloch, and C. Gross, \emph{{Crystallization in Ising quantum
  magnets.}}, Science (80-. ). 347 (2015), p. 1455,
  \urlprefix\url{http://arxiv.org/abs/1404.0980}.

\bibitem{Wiese2014}
U.J. Wiese, \emph{{Towards quantum simulating QCD}}, Nucl. Phys. A 931 (2014),
  pp. 246--256,
  \urlprefix\url{http://linkinghub.elsevier.com/retrieve/pii/S0375947414004849}.

\bibitem{Zohar2015a}
E. Zohar, J.I. Cirac, and B. Reznik, \emph{{Quantum Simulations of Lattice
  Gauge Theories using Ultracold Atoms in Optical Lattices}}
  \urlprefix\url{http://arxiv.org/abs/1503.02312}.

\bibitem{Glaetzle2015}
A.W. Glaetzle, M. Dalmonte, R. Nath, C. Gross, I. Bloch, and P. Zoller,
  \emph{{Designing Frustrated Quantum Magnets with Laser-Dressed Rydberg
  Atoms}}, Phys. Rev. Lett. 114 (2015), p. 173002,
  \urlprefix\url{http://link.aps.org/doi/10.1103/PhysRevLett.114.173002}.

\bibitem{Egusquiza2015}
A. Mezzacapo, E. Rico, C. Sab{\'\i}n, I.L. Egusquiza, L. Lamata, and E. Solano,
  \emph{Non-abelian $su(2)$ lattice gauge theories in superconducting
  circuits}, Phys. Rev. Lett. 115 (2015), p. 240502,
  \urlprefix\url{http://arxiv.org/abs/1505.04720}.

\bibitem{Barrett2013}
S. Barrett, K. Hammerer, S. Harrison, T.E. Northup, and T.J. Osborne,
  \emph{{Simulating quantum fields with cavity QED}}, Phys. Rev. Lett. 110
  (2013), pp. 1--6.

\bibitem{Hauke2013}
P. Hauke, D. Marcos, M. Dalmonte, and P. Zoller, \emph{{Quantum Simulation of a
  Lattice Schwinger Model in a Chain of Trapped Ions}}, Phys. Rev. X 3 (2013),
  p. 041018, \urlprefix\url{http://link.aps.org/doi/10.1103/PhysRevX.3.041018}.

\bibitem{Banerjee2013}
D. Banerjee, M. B{\"{o}}gli, M. Dalmonte, E. Rico, P. Stebler, U.J. Wiese, and
  P. Zoller, \emph{{Atomic Quantum Simulation of U(N) and SU(N) Non-Abelian
  Lattice Gauge Theories}}, Phys. Rev. Lett. 110 (2013), p. 125303,
  \urlprefix\url{http://link.aps.org/doi/10.1103/PhysRevLett.110.125303}.

\bibitem{Rico2014}
E. Rico, T. Pichler, M. Dalmonte, P. Zoller, and S. Montangero, \emph{{Tensor
  Networks for Lattice Gauge Theories and Atomic Quantum Simulation}}, Phys.
  Rev. Lett. 112 (2014), p. 201601,
  \urlprefix\url{http://link.aps.org/doi/10.1103/PhysRevLett.112.201601}.

\bibitem{Pichler2015}
T. Pichler, M. Dalmonte, E. Rico, P. Zoller, and S. Montangero,
  \emph{{Real-time Dynamics in U(1) Lattice Gauge Theories with Tensor
  Networks}}, arXiv:1505.04440 \urlprefix\url{http://arxiv.org/abs/1505.04440}.

\bibitem{Tagliacozzo2014}
L. Tagliacozzo, A. Celi, and M. Lewenstein, \emph{{Tensor Networks for Lattice
  Gauge Theories with Continuous Groups}}, Phys. Rev. X 4 (2014), p. 041024,
  \urlprefix\url{http://link.aps.org/doi/10.1103/PhysRevX.4.041024}.

\bibitem{Banuls}
M. Ba{\~{n}}uls, K. Cichy, J. Cirac, and K. Jansen, \emph{{The mass spectrum of
  the Schwinger model with matrix product states}}, J. High Energy Phys. 2013
  (2013), p. 158, \urlprefix\url{http://arxiv.org/abs/1305.3765
  http://dx.doi.org/10.1007/JHEP11(2013)158
  http://link.springer.com/10.1007/JHEP11(2013)158}.

\bibitem{Banuls2015}
M.C. Ba{\~{n}}uls, K. Cichy, J.I. Cirac, K. Jansen, and H. Saito,
  \emph{{Thermal evolution of the Schwinger model with matrix product
  operators}}, Phys. Rev. D 92 (2015), p. 034519,
  \urlprefix\url{http://arxiv.org/abs/1505.00279
  http://dx.doi.org/10.1103/PhysRevD.92.034519
  http://link.aps.org/doi/10.1103/PhysRevD.92.034519}.

\bibitem{Buyens2014}
B. Buyens, J. Haegeman, K. {Van Acoleyen}, H. Verschelde, and F. Verstraete,
  \emph{{Matrix Product States for Gauge Field Theories}}, Phys. Rev. Lett. 113
  (2014), p. 091601,
  \urlprefix\url{http://link.aps.org/doi/10.1103/PhysRevLett.113.091601}.

\bibitem{Lacroix2010}
C. Lacroix, P. Mendels, and F. Mila (eds.), \emph{Introduction to Frustrated
  Magnetism}, Springer Series in Solid-State Sciences Vol. 164, 2010.

\bibitem{Nayak2008}
C. Nayak, A. Stern, M. Freedman, and S. {Das Sarma}, \emph{{Non-Abelian anyons
  and topological quantum computation}}, Rev. Mod. Phys. 80 (2008), pp.
  1083--1159,
  \urlprefix\url{http://link.aps.org/doi/10.1103/RevModPhys.80.1083}.

\bibitem{Kitaev2003}
a.Y. Kitaev, \emph{{Fault-tolerant quantum computation by anyons}}, Ann. Phys.
  (N. Y). 303 (2003), pp. 2--30.

\bibitem{VanWezel2007}
J. van  Wezel and J. van~den  Brink, \emph{{Spontaneous symmetry breaking in
  quantum mechanics}}, Am. J. Phys. 75 (2007), p. 635,
  \urlprefix\url{http://scitation.aip.org/content/aapt/journal/ajp/75/7/10.1119/1.2730839}.

\bibitem{Goldstein1980}
H. Goldstein, \emph{{Classical Mechanics}}, Addison-Wesley, Reading, MA, 1980.

\bibitem{Greiner2002}
M. Greiner, O. Mandel, T. Esslinger, T.W. H{\"{a}}nsch, and I. Bloch,
  \emph{{Quantum phase transition from a superfluid to a Mott insulator in a
  gas of ultracold atoms}}, Nature 415 (2002), pp. 39--44,
  \urlprefix\url{http://dx.doi.org/10.1038/415039a}.

\bibitem{Wilson74}
K.G. Wilson, \emph{Confinement of quarks}, Phys. Rev. D 10 (1974), pp.
  2445--2459.

\bibitem{Montvay1994}
I. Montvay and G. Muenster, \emph{Quantum Fields on a lattice}, Cambridge Univ.
  Press, Cambridge, 1994.

\bibitem{Creutz1997}
M. Creutz, \emph{Quarks, gluons and lattices}, Cambridge University Press,
  Cambridge, 1997.

\bibitem{DeGrand2006}
T. DeGrand and C. DeTar, \emph{Lattice Methods for Quantum Chromodynamics},
  World Scientific, 2006.

\bibitem{Gattringer2010}
C. Gattringer and C.B. Lang, \emph{Quantum Chromodynamics on the Lattice},
  Springer-Verlag, 2010.

\bibitem{Kogut1975}
J. Kogut and L. Susskind, \emph{{Hamiltonina formulation of Wilson's lattice
  guage theoreis}}, Phys. Rev. D 11 (1975).

\bibitem{Kogut1979}
J. Kogut, \emph{{An introduction to lattice gauge theory and spin systems}},
  Rev. Mod. Phys.  (1979),
  \urlprefix\url{http://rmp.aps.org/abstract/RMP/v51/i4/p659{\_}1}.

\bibitem{Horn1981}
D. Horn, \emph{Finite matrix models with continuous local gauge invariance},
  Physics Letters B 100 (1981), p. 149.

\bibitem{Orland1990}
P. Orland and D. Rohrlich, \emph{Lattice gauge magnets: local isospin from
  spin.}, Nucl. Phys. B 338 (1990), p. 647.

\bibitem{Chandrasekharan1997}
S. Chandrasekharan and U.J. Wiese, \emph{{Quantum link models : A discrete
  approach to gauge theories}}, Nucl. Phys. B 492 (1997), pp. 455--471,

\bibitem{wegner71}
F. Wegner, \emph{Duality in generalized ising models and phase transitions
  without local order parameters}, J. Math. Phys. 10 (1971), p. 2259.

\bibitem{Mathur:2005cr}
M. Mathur, \emph{Harmonic oscillator prepotentials in su(2) lattice gauge
  theory}, J.Phys. A38 (2005), pp. 10015--10026,
  \urlprefix\url{http://arxiv.org/abs/hep-lat/0403029}.

\bibitem{Anishetty:2010dq}
R. Anishetty, M. Mathur, and I. Raychowdhury, \emph{Prepotential formulation of
  su(3) lattice gauge theory}, J.Phys.A 43 (2010), p. 035403,
  \urlprefix\url{http://arxiv.org/abs/0909.2394}.

\bibitem{Zohar2015}
E. Zohar and M. Burrello, \emph{{Formulation of lattice gauge theories for
  quantum simulations}}, Phys. Rev. D 91 (2015), p. 054506,
  \urlprefix\url{http://arxiv.org/abs/1409.3085
  http://dx.doi.org/10.1103/PhysRevD.91.054506
  http://link.aps.org/doi/10.1103/PhysRevD.91.054506}.

\bibitem{Brower1999}
R. Brower, S. Chandrasekharan, and U.J. Wiese, \emph{{QCD} as a quantum link
  model}, Phys. Rev. D 60 (1999), p. 094502.

\bibitem{baxterbook}
R.J. Baxter, \emph{Exactly Solved Models in Statistical Mechanics}, Dover
  Publications, 2007.

\bibitem{Wiese:2013kk}
U.J. Wiese, \emph{{Ultracold quantum gases and lattice systems: quantum
  simulation of lattice gauge theories}}, Annalen der Physik 525 (2013), pp.
  777--796.

\bibitem{Banerjee2012}
D. Banerjee, M. Dalmonte, M. M{\"{u}}ller, E. Rico, P. Stebler, U.J. Wiese, and
  P. Zoller, \emph{{Atomic Quantum Simulation of Dynamical Gauge Fields Coupled
  to Fermionic Matter: From String Breaking to Evolution after a Quench}},
  Phys. Rev. Lett. 109 (2012), pp. 1--5,
  \urlprefix\url{http://link.aps.org/doi/10.1103/PhysRevLett.109.175302}.

\bibitem{BernevigBook}
B.A. Bernevig, \emph{Topological Insulators and Topological Superconductors},
  Princeton University Press, 2013.

\bibitem{Shannon:2004nr}
N. Shannon, G. Misguich, and K. Penc, \emph{Cyclic exchange, isolated states
  and spinon deconfinement in an xxz heisenberg model on the checkerboard
  lattice}, Phys. Rev. B 69 (2004), p. 220403(R),
  \urlprefix\url{http://arxiv.org/abs/cond-mat/0403729}.

\bibitem{Banerjee:2013gf}
D. Banerjee, F.J. Jiang, P. Widmer, and U.J. Wiese, \emph{The (2+1)-d u(1)
  quantum link model masquerading as deconfined criticality}  (2013),
  \urlprefix\url{http://arxiv.org/abs/1303.6858}.

\bibitem{Beard1998775}
B. Beard, R. Brower, S. Chandrasekharan, D. Chen, A. Tsapalis, and U.J. Wiese,
  \emph{D-theory: field theory via dimensional reduction of discrete
  variables}, Nuclear Physics B - Proceedings Supplements 63 (1998), pp. 775 --
  789,
  \urlprefix\url{http://www.sciencedirect.com/science/article/pii/S0920563297009006},
  proceedings of the \{XVth\} International Symposium on Lattice Field Theory.

\bibitem{Eisert2010}
J. Eisert, M. Cramer, and M.B. Plenio, \emph{{<i>Colloquium</i> : Area laws for
  the entanglement entropy}}, Rev. Mod. Phys. 82 (2010), pp. 277--306,
  \urlprefix\url{http://link.aps.org/doi/10.1103/RevModPhys.82.277}.

\bibitem{Lubasch2014}
M. Lubasch, J.I. Cirac, and M.C. Ba{\~{n}}uls, \emph{{Algorithms for finite
  Projected Entangled Pair States}}  (2014), p.~18,
  \urlprefix\url{http://arxiv.org/abs/1405.3259}.

\bibitem{Verstraete2004}
F. Verstraete, J.J. Garc{\'{\i}}a-Ripoll, and J.I. Cirac, \emph{{Matrix Product
  Density Operators: Simulation of Finite-Temperature and Dissipative
  Systems}}, Phys. Rev. Lett. 93 (2004), p. 207204,
  \urlprefix\url{http://link.aps.org/doi/10.1103/PhysRevLett.93.207204}.

\bibitem{Kliesch2014b}
M. Kliesch, D. Gross, and J. Eisert, \emph{{Matrix-Product Operators and
  States: NP-Hardness and Undecidability}}, Phys. Rev. Lett. 113 (2014), p.
  160503,
  \urlprefix\url{http://link.aps.org/doi/10.1103/PhysRevLett.113.160503}.

\bibitem{Tagliacozzo2009}
L. Tagliacozzo, G. Evenbly, and G. Vidal, \emph{{Simulation of two-dimensional
  quantum systems using a tree tensor network that exploits the entropic area
  law}}, Phys. Rev. B 80 (2009), p. 235127,
  \urlprefix\url{http://link.aps.org/doi/10.1103/PhysRevB.80.235127}.

\bibitem{Evenbly2012}
G. Evenbly and G. Vidal, \emph{{Class of Highly Entangled Many-Body States that
  can be Efficiently Simulated}}, Phys. Rev. Lett. 112 (2014), p. 240502,
  \urlprefix\url{http://arxiv.org/abs/1210.1895
  http://dx.doi.org/10.1103/PhysRevLett.112.240502
  http://link.aps.org/doi/10.1103/PhysRevLett.112.240502}.

\bibitem{Gerster2014b}
M. Gerster, P. Silvi, M. Rizzi, R. Fazio, T. Calarco, and S. Montangero,
  \emph{{Unconstrained tree tensor network: An adaptive gauge picture for
  enhanced performance}}, Phys. Rev. B 90 (2014), p. 125154,
  \urlprefix\url{http://arxiv.org/abs/1406.2666
  http://link.aps.org/doi/10.1103/PhysRevB.90.125154}.

\bibitem{Vidal2007}
G. Vidal, \emph{{Entanglement Renormalization}} 220405 (2007), pp. 1--4.

\bibitem{Cuevas2013}
G.D. las  Cuevas, N. Schuch, D. P{\'{e}}rez-Garc{\'{\i}}a, and J. {Ignacio
  Cirac}, \emph{{Purifications of multipartite states: limitations and
  constructive methods}}, New J. Phys. 15 (2013), pp. 123021--123026.

\bibitem{Werner2014}
A.H. Werner, D. Jaschke, P. Silvi, T. Calarco, J. Eisert, and S. Montangero,
  \emph{{A positive tensor network approach for simulating open quantum
  many-body systems}}, arXiv:1412.5746
  \urlprefix\url{http://arxiv.org/abs/1412.5746}.

\bibitem{Daley2014}
A.J. Daley, \emph{{Quantum trajectories and open many-body quantum systems}},
  Adv. Phys. 63 (2014), pp. 77--149,
  \urlprefix\url{http://www.tandfonline.com/doi/abs/10.1080/00018732.2014.933502}.

\bibitem{Bonnes2014}
L. Bonnes and A.M. L{\"{a}}uchli, \emph{{Superoperators vs. Trajectories for
  Matrix Product State Simulations of Open Quantum System: A Case Study}},
  arXiv Prepr. arXiv1411.4831 2 (2014), pp. 1--10,
  \urlprefix\url{http://arxiv.org/abs/1411.4831}.

\bibitem{Silvi2014}
P. Silvi, E. Rico, T. Calarco, and S. Montangero, \emph{{Lattice gauge tensor
  networks}}, New J. Phys. 16 (2014), p. 103015,
  \urlprefix\url{http://arxiv.org/abs/1404.7439
  http://stacks.iop.org/1367-2630/16/i=10/a=103015?key=crossref.b1a846ff30279981436f10cdcaa0303c}.

\bibitem{Notarnicola2015a}
S. Notarnicola, E. Ercolessi, P. Facchi, G. Marmo, S. Pascazio, and F.V. Pepe,
  \emph{{Discrete Abelian Gauge Theories for Quantum Simulations of QED}}, J.
  Phys. A Math. Theor. 48 (2015), pp. 1--13,
  \urlprefix\url{http://arxiv.org/abs/1503.04340$\backslash$nhttp://dx.doi.org/10.1088/1751-8113/48/30/30FT01}.

\bibitem{Osborne}
T. Osborne, \emph{{Lattice Gauge Theories and Tensor Networks}},
  \urlprefix\url{https://github.com/tobiasosborne/Lattice‑gauge‑theory‑and‑tensor‑networks}.

\bibitem{Haegeman2015}
J. Haegeman, K. {Van Acoleyen}, N. Schuch, J.I. Cirac, and F. Verstraete,
  \emph{{Gauging Quantum States: From Global to Local Symmetries in Many-Body
  Systems}}, Phys. Rev. X 5 (2015), p. 011024,
  \urlprefix\url{http://link.aps.org/doi/10.1103/PhysRevX.5.011024}.

\bibitem{Tagliacozzo2011}
L. Tagliacozzo and G. Vidal, \emph{{Entanglement renormalization and gauge
  symmetry}}, Phys. Rev. B 83 (2011), p. 115127,
  \urlprefix\url{http://link.aps.org/doi/10.1103/PhysRevB.83.115127}.

\bibitem{Sugihara2005}
T. Sugihara, \emph{{Matrix product representation of gauge invariant states in
  a Bbb Z 2 lattice gauge theory}}, J. High Energy Phys. 2005 (2005), pp.
  022--022, \urlprefix\url{http://iopscience.iop.org/1126-6708/2005/07/022
  http://stacks.iop.org/1126-6708/2005/i=07/a=022?key=crossref.033dfd8a23cc493d05da616fdaec0de7}.

\bibitem{Dusuel2011}
S. Dusuel, M. Kamfor, R. Or{\'{u}}s, K.P. Schmidt, and J. Vidal,
  \emph{{Robustness of a perturbed topological phase}}, Phys. Rev. Lett. 106
  (2011), pp. 1--4.

\bibitem{Kuhn2014}
S. K{\"{u}}hn, J.I. Cirac, and M.C. Ba{\~{n}}uls, \emph{{Quantum simulation of
  the Schwinger model: A study of feasibility}}, Phys. Rev. A 90 (2014), p.
  042305, \urlprefix\url{http://link.aps.org/doi/10.1103/PhysRevA.90.042305}.

\bibitem{Kuhn2015}
S. K{\"{u}}hn, E. Zohar, J.I. Cirac, and M.C. Ba{\~{n}}uls, \emph{{Non-Abelian
  string breaking phenomena with matrix product states}}, J. High Energy Phys.
  2015 (2015), p. 130,
  \urlprefix\url{http://link.springer.com/10.1007/JHEP07(2015)130}.

\bibitem{Byrnes2002}
T. Byrnes, P. Sriganesh, R. Bursill, and C. Hamer, \emph{{Density matrix
  renormalisation group approach to the massive Schwinger model}}, Phys. Rev. D
  66 (2002), p. 013002,
  \urlprefix\url{http://link.aps.org/doi/10.1103/PhysRevD.66.013002$\backslash$nhttp://arxiv.org/abs/hep-lat/0201007$\backslash$nhttp://linkinghub.elsevier.com/retrieve/pii/S0920563202014160
  http://linkinghub.elsevier.com/retrieve/pii/S0920563202014160}.

\bibitem{Banuls:2013qf}
M.C. Ba{\~n}uls, K. Cichy, J.I. Cirac, K. Jansen, and H. Saito, \emph{Matrix
  product states for lattice field theories}, PoS(LATTICE 2013)332  (2013),
  \urlprefix\url{http://arxiv.org/abs/1310.4118}.

\bibitem{Saito:2015kq}
H. Saito, M.C. Ba{\~n}uls, K. Cichy, J.I. Cirac, and K. Jansen, \emph{Thermal
  evolution of the one-flavour schwinger model using matrix product states}
  (2015), \urlprefix\url{http://arxiv.org/abs/1511.00794}.

\bibitem{Zohar:2015yq}
E. Zohar, M. Burrello, T.B. Wahl, and J.I. Cirac, \emph{Fermionic projected
  entangled pair states and local u(1) gauge theories}, Annals of Physics
  (2015), pp. 385-439  (2015), \urlprefix\url{http://arxiv.org/abs/1507.08837}.

\bibitem{Zohar:2015bf}
E. Zohar and M. Burrello, \emph{Building projected entangled pair states with a
  local gauge symmetry}  (2015),
  \urlprefix\url{http://arxiv.org/abs/1511.08426}.

\bibitem{wenbook}
X.G. Wen, \emph{Quantum Field Theory of Many-Body Systems}, Oxford University
  Press, 2004.

\bibitem{fradkinbook}
E. Fradkin, \emph{{Field Theories of Condensed Matter Systems}}, Cambridge
  University Press, 2013.

\bibitem{Diehl:2008xy}
S. Diehl, A. Micheli, A. Kantian, B. Kraus, H. B{\"u}chler, and P. Zoller,
  \emph{Quantum states and phases in driven open quantum systems with cold
  atoms}, Nature Physics 4 (2008), p. 878,
  \urlprefix\url{http://arxiv.org/abs/0803.1482}.

\bibitem{Verstraete:2008sf}
F. Verstraete, M.M. Wolf, and J.I. Cirac, \emph{Quantum computation, quantum
  state engineering, and quantum phase transitions driven by dissipation},
  Nature Physics 5, 633 - 636 (2009)  (2008),
  \urlprefix\url{http://arxiv.org/abs/0803.1447}.

\bibitem{Kantian:2009rw}
A. Kantian, M. Dalmonte, S. Diehl, W. Hofstetter, P. Zoller, and A.J. Daley,
  \emph{An atomic colour superfluid via three-body loss}, Phys. Rev. Lett. 103
  (2009), p. 240401, \urlprefix\url{http://arxiv.org/abs/0908.3235}.

\bibitem{Daley:2009qr}
A.J. Daley, J.M. Taylor, S. Diehl, M. Baranov, and P. Zoller, \emph{Atomic
  three-body loss as a dynamical three-body interaction}, Phys. Rev. Lett. 102
  (2009), p. 040402, \urlprefix\url{http://arxiv.org/abs/0810.5153}.

\bibitem{Syassen:2008it}
N. Syassen, D.M. Bauer, M. Lettner, T. Volz, D. Dietze, J.J. Garcia-Ripoll,
  J.I. Cirac, G. Rempe, and S. D{\"u}rr, \emph{Strong dissipation inhibits
  losses and induces correlations in cold molecular gases}, Science 320 (2008),
  p. 1329, \urlprefix\url{http://arxiv.org/abs/0806.4310}.

\bibitem{Garcia-Ripoll:2009ek}
J.J. Garcia-Ripoll, S. D{\"u}rr, N. Syassen, D.M. Bauer, M. Lettner, G. Rempe,
  and J.I. Cirac, \emph{Dissipation induced tonks-girardeau gas in an optical
  lattice}, New J. Phys. 11 (2009), p. 013053,
  \urlprefix\url{http://arxiv.org/abs/0809.3679}.

\bibitem{Muschik:2012fv}
C.A. Muschik, H. Krauter, K. Jensen, J.M. Petersen, J.I. Cirac, and E.S.
  Polzik, \emph{Robust entanglement generation by reservoir engineering}, J.
  Phys. B: At. Mol. Opt. Phys. 45 (2012), p. 124021,
  \urlprefix\url{http://arxiv.org/abs/1203.4785}.

\bibitem{Facchi:2000bs}
P. Facchi, V. Gorini, G. Marmo, S. Pascazio, and E. Sudarshan, \emph{Quantum
  zeno dynamics}, Phys. Lett. A, 275 (2000), p.~12,
  \urlprefix\url{http://arxiv.org/abs/quant-ph/0004040}.

\bibitem{Facchi:2002ij}
P. Facchi and S. Pascazio, \emph{Quantum zeno subspaces}, Phys. Rev. Lett. 89
  (2002), p. 080401, \urlprefix\url{http://arxiv.org/abs/quant-ph/0201115}.

\bibitem{Stannigel:2014bf}
K. Stannigel, P. Hauke, D. Marcos, M. Hafezi, S. Diehl, M. Dalmonte, and P.
  Zoller, \emph{Constrained dynamics via the zeno effect in quantum simulation:
  Implementing non-abelian lattice gauge theories with cold atoms}, Phys. Rev.
  Lett. 112 (2014), p. 120406, \urlprefix\url{http://arxiv.org/abs/1308.0528}.

\bibitem{gardiner_book}
C. Gardiner and P. Zoller, \emph{The Quantum World of Ultra-Cold Atoms and
  Light Book I: Foundations of Quantum Optics}, Imperial College Press, 2014.

\bibitem{Zohar:2013qf}
E. Zohar, J.I. Cirac, and B. Reznik, \emph{Quantum simulations of gauge
  theories with ultracold atoms: local gauge invariance from angular momentum
  conservation}, Phys. Rev. A 88 (2013), p. 023617,
  \urlprefix\url{http://arxiv.org/abs/1303.5040}.

\bibitem{Zohar:2013kb}
E. Zohar, J.I. Cirac, and B. Reznik, \emph{A cold-atom quantum simulator for
  su(2) yang-mills lattice gauge theory}, Phys. Rev. Lett. 110 (2013), p.
  125304, \urlprefix\url{http://arxiv.org/abs/1211.2241}.

\bibitem{Kasper:2015kc}
V. Kasper, F. Hebenstreit, M. Oberthaler, and J. Berges, \emph{Schwinger pair
  production with ultracold atoms}  (2015),
  \urlprefix\url{http://arxiv.org/abs/1506.01238}.

\bibitem{lloyd1996universal}
S. Lloyd, \emph{Universal quantum simulators}, Science  (1996), p. 1073.

\bibitem{Tagliacozzo:2012kq}
L. Tagliacozzo, A. Celi, A. Zamora, and M. Lewenstein, \emph{Optical abelian
  lattice gauge theories}, Annals of Physics (2013), pp. 160-191  (2012),
  \urlprefix\url{http://arxiv.org/abs/1205.0496}.

\bibitem{Weimer:2010xw}
H. Weimer, M. M{\"u}ller, I. Lesanovsky, P. Zoller, and H.P. B{\"u}chler,
  \emph{A rydberg quantum simulator}, Nature Phys. 6 (2010), pp. 382--388,
  \urlprefix\url{http://arxiv.org/abs/0907.1657}.

\bibitem{greiner2002quantum}
M. Greiner, O. Mandel, T. Esslinger, T.W. H{\"a}nsch, and I. Bloch,
  \emph{Quantum phase transition from a superfluid to a mott insulator in a gas
  of ultracold atoms}, nature 415 (2002), pp. 39--44.

\bibitem{jaksch1998cold}
D. Jaksch, C. Bruder, J.I. Cirac, C.W. Gardiner, and P. Zoller, \emph{Cold
  bosonic atoms in optical lattices}, Physical Review Letters 81 (1998), p.
  3108.

\bibitem{jordens2008mott}
R. J{\"o}rdens, N. Strohmaier, K. G{\"u}nter, H. Moritz, and T. Esslinger,
  \emph{A mott insulator of fermionic atoms in an optical lattice}, Nature 455
  (2008), pp. 204--207.

\bibitem{schneider2008metallic}
U. Schneider, L. Hackerm{\"u}ller, S. Will, T. Best, I. Bloch, T. Costi, R.
  Helmes, D. Rasch, and A. Rosch, \emph{Metallic and insulating phases of
  repulsively interacting fermions in a 3d optical lattice}, Science 322
  (2008), pp. 1520--1525.

\bibitem{Fallani14}
G. Pagano, M. Mancini, G. Cappellini, P. Lombardi, F. Schafer, H. Hu, X.J. Liu,
  J. Catani, C. Sias, M. Inguscio, and L. Fallani, \emph{A one-dimensional
  liquid of fermions with tunable spin}, Nat Phys 10 (2014), pp. 198--201.

\bibitem{Taie:2012ys}
S. Taie, R. Yamazaki, S. Sugawa, and Y. Takahashi, \emph{An su(n) mott
  insulator of an atomic fermi gas realized by large-spin pomeranchuk cooling},
  Nat. Phys. 8 (2012), p. 825.

\bibitem{Hofrichter:2015fv}
C. Hofrichter, L. Riegger, F. Scazza, M. H{\"o}fer, D.R. Fernandes, I. Bloch,
  and S. F{\"o}lling, \emph{Direct probing of the mott crossover in the su($n$)
  fermi-hubbard model}  (2015),
  \urlprefix\url{http://arxiv.org/abs/1511.07287}.

\bibitem{Paz:2013dn}
A. de  Paz", \emph{Nonequilibrium quantum magnetism in a dipolar lattice gas},
  Physical Review Letters 111 (2013).

\bibitem{Yan:2013bh}
B. Yan, S.A. Moses, B. Gadway, J.P. Covey, K.R.A. Hazzard, A.M. Rey, D.S. Jin,
  and J. Ye, \emph{Realizing a lattice spin model with polar molecules}, Nature
  501 521 (2013)  (2013), \urlprefix\url{http://arxiv.org/abs/1305.5598}.

\bibitem{Schaus:2015it}
P. Schau{\ss}, J. Zeiher, T. Fukuhara, S. Hild, M. Cheneau, T. Macr{\`\i}, T.
  Pohl, I. Bloch, and C. Gross, \emph{Dynamical crystallization in a
  low-dimensional rydberg gas}, Science 347 (2015), p. 1455,
  \urlprefix\url{http://arxiv.org/abs/1404.0980}.

\bibitem{Atala:2014th}
M. Atala, M. Aidelsburger, M. Lohse, J.T. Barreiro, B. Paredes, and I. Bloch,
  \emph{Observation of the meissner effect with ultracold atoms in bosonic
  ladders}, Nature Physics 10 (2014), pp. 588--593,
  \urlprefix\url{http://arxiv.org/abs/1402.0819}.

\bibitem{Mancini25092015}
M. Mancini, G. Pagano, G. Cappellini, L. Livi, M. Rider, J. Catani, C. Sias, P.
  Zoller, M. Inguscio, M. Dalmonte, and L. Fallani, \emph{Observation of chiral
  edge states with neutral fermions in synthetic hall ribbons}, Science 349
  (2015), pp. 1510--1513,
  \urlprefix\url{http://www.sciencemag.org/content/349/6255/1510.abstract}.

\bibitem{Miyake:2013dz}
H. Miyake, G.A. Siviloglou, C.J. Kennedy, W.C. Burton, and W. Ketterle,
  \emph{Realizing the harper hamiltonian with laser-assisted tunneling in
  optical lattices}, Phys. Rev. Lett. 111 (2013), p. 185302,
  \urlprefix\url{http://arxiv.org/abs/1308.1431}.

\bibitem{Struck:2013yf}
J. Struck, M. Weinberg, C. {\"O}lschl{\"a}ger, P. Windpassinger, J. Simonet, K.
  Sengstock, R. H{\"o}ppner, P. Hauke, A. Eckardt, M. Lewenstein, and L.
  Mathey, \emph{Engineering ising-xy spin models in a triangular lattice via
  tunable artificial gauge fields}  (2013),
  \urlprefix\url{http://arxiv.org/abs/1304.5520}.

\bibitem{Stuhl:2015rw}
B.K. Stuhl, H.I. Lu, L.M. Aycock, D. Genkina, and I.B. Spielman,
  \emph{Visualizing edge states with an atomic bose gas in the quantum hall
  regime}, Science 349 (2015), pp. 1514--1518,
  \urlprefix\url{http://arxiv.org/abs/1502.02496}.

\bibitem{Jaksch2005}
D. Jaksch and P. Zoller, \emph{{The cold atom Hubbard toolbox}}, Ann. Phys. (N.
  Y). 315 (2005), pp. 52--79.

\bibitem{Chin:2010fe}
C. Chin, R. Grimm, P. Julienne, and E. Tiesinga, \emph{Feshbach resonances in
  ultracold gases}, Rev. Mod. Phys. 82 (2010), pp. 1225--1286,
  \urlprefix\url{http://arxiv.org/abs/0812.1496}.

\bibitem{Baranov:2012wo}
M.A. Baranov, M. Dalmonte, G. Pupillo, and P. Zoller, \emph{Condensed matter
  theory of dipolar quantum gases}, Chemical Reviews 112 (2012), p. 5012,
  \urlprefix\url{http://arxiv.org/abs/1207.1914}.

\bibitem{Lahaye:2009by}
T. Lahaye, C. Menotti, L. Santos, M. Lewenstein, and T. Pfau, \emph{The physics
  of dipolar bosonic quantum gases}, Rep. Prog. Phys. 72 (2009), p. 126401,
  \urlprefix\url{http://arxiv.org/abs/0905.0386}.

\bibitem{Saffman:2010bq}
M. Saffman, T.G. Walker, and K. Molmer, \emph{Quantum information with rydberg
  atoms}, Rev. Mod. Phys. 82 (2010), p. 2313,
  \urlprefix\url{http://arxiv.org/abs/0909.4777}.

\bibitem{Kapit:2010kh}
E. Kapit and E.J. Mueller, \emph{Optical lattice hamiltonians for relativistic
  quantum electrodynamics}  (2010),
  \urlprefix\url{http://arxiv.org/abs/1011.4021}.

\bibitem{Zohar:2013om}
E. Zohar, J.I. Cirac, and B. Reznik, \emph{Simulating 2+1d lattice qed with
  dynamical matter using ultracold atoms}, Phys. Rev. Lett. 110 (2013), p.
  055302, \urlprefix\url{http://arxiv.org/abs/1208.4299}.

\bibitem{Bazavov:2015zh}
A. Bazavov, Y. Meurice, S.W. Tsai, J. Unmuth-Yockey, and J. Zhang,
  \emph{Gauge-invariant implementation of the abelian higgs model on optical
  lattices}, Phys. Rev. D 92 (2015), p. 076003,
  \urlprefix\url{http://arxiv.org/abs/1503.08354}.

\bibitem{Laflamme:2015oe}
C. Laflamme, W. Evans, M. Dalmonte, U. Gerber, H. Mej{\'\i}a-D{\'\i}az, W.
  Bietenholz, U.J. Wiese, and P. Zoller, \emph{Cp(n-1) quantum field theories
  with alkaline-earth atoms in optical lattices}  (2015),
  \urlprefix\url{http://arxiv.org/abs/1507.06788}.

\bibitem{Laflamme:2015sy}
C. Laflamme, W. Evans, M. Dalmonte, U. Gerber, H. Mej{\'\i}a-D{\'\i}az, W.
  Bietenholz, U.J. Wiese, and P. Zoller, \emph{Proposal for the quantum
  simulation of the cp(2) model on optical lattices}  (2015),
  \urlprefix\url{http://arxiv.org/abs/1510.08492}.

\bibitem{ALV}
A. d'Adda, M. L{\"u}scher, and P. Di~Vecchia, \emph{A 1n expandable series of
  non-linear $\sigma$ models with instantons}, Nuclear Physics B 146 (1978),
  pp. 63--76.

\bibitem{Zohar:2012kl}
E. Zohar", \emph{Simulating compact quantum electrodynamics with ultracold
  atoms: Probing confinement and nonperturbative effects}, Physical Review
  Letters 109 (2012).

\bibitem{Tewari:2006ss}
S. Tewari, V.W. Scarola, T. Senthil, and S.D. Sarma, \emph{Emergence of
  artificial photons in an optical lattice}, Phys. Rev. Lett. 97 (2006), p.
  200401, \urlprefix\url{http://arxiv.org/abs/cond-mat/0605154}.

\bibitem{Bijnen:2015bx}
R.M.W. van  Bijnen and T. Pohl, \emph{Quantum magnetism and topological
  ordering via enhanced rydberg-dressing near f\"orster-resonances}, Phys. Rev.
  Lett. 114 (2015), p. 243002, \urlprefix\url{http://arxiv.org/abs/1411.3118}.

\bibitem{gerritsma2010quantum}
R. Gerritsma, G. Kirchmair, F. Z{\"a}hringer, E. Solano, R. Blatt, and C. Roos,
  \emph{Quantum simulation of the dirac equation}, Nature 463 (2010), pp.
  68--71.

\bibitem{Porras:2004km}
D. Porras and J. Cirac, \emph{Effective quantum spin systems with ion traps},
  Phys. Rev. Lett. 92 (2004), p. 207901,
  \urlprefix\url{http://arxiv.org/abs/quant-ph/0401102}.

\bibitem{schneider2011many}
C. Schneider, D. Porras, and T. Schaetz, \emph{Many-body physics with trapped
  ions}, arXiv preprint arXiv:1106.2597  (2011).

\bibitem{Nath:2015to}
R. Nath, M. Dalmonte, A.W. Glaetzle, P. Zoller, F. Schmidt-Kaler, and R.
  Gerritsma, \emph{Hexagonal plaquette spin-spin interactions and quantum
  magnetism in a two-dimensional ion crystal}, New J. Phys. 17 (2015), p.
  065018, \urlprefix\url{http://arxiv.org/abs/1504.01474}.

\bibitem{Balents:2002yi}
L. Balents", \emph{Fractionalization in an easy-axis kagome antiferromagnet},
  Physical Review B 65 (2002).

\bibitem{Isakov:2012wl}
S.V. Isakov, R.G. Melko, and M.B. Hastings, \emph{Universal signatures of
  fractionalized quantum critical points}, Science 335 (2012), pp. 193--195,
  \urlprefix\url{http://arxiv.org/abs/1108.2055}.

\bibitem{Houck:2012}
A.A. Houck, H.E. T{\"u}reci, and J. Koch, \emph{{On-chip quantum simulation
  with superconducting circuits}}, Nature Physics 8 (2012), pp. 292--299.

\bibitem{Viehmann2013a}
O. Viehmann, J. von  Delft, and F. Marquardt, \emph{{Observing the
  Nonequilibrium Dynamics of the Quantum Transverse-Field Ising Chain in
  Circuit QED}}, Phys. Rev. Lett. 110 (2013), p. 030601.

\bibitem{Mezzacapo:2014bs}
A. Mezzacapo, L. Lamata, S. Filipp, and E. Solano, \emph{Many-body interactions
  with tunable-coupling transmon qubits}, Phys. Rev. Lett. 113 (2014), p.
  050501.

\bibitem{Naether:2014dz}
U. Naether, F. Quijandr{\'\i}a, J.J. Garc{\'\i}a-Ripoll, and D. Zueco,
  \emph{Stationary discrete solitons in circuit qed}  (2014).

\bibitem{Doucot:2004rm}
B. Doucot, L. Ioffe, and J. Vidal, \emph{Discrete non-abelian gauge theories in
  two-dimensional lattices and their realizations in josephson-junction
  arrays}, Phys.Rev.B 69 (2004), p. 214501,
  \urlprefix\url{http://arxiv.org/abs/cond-mat/0302104}.

\bibitem{Marcos2013}
D. Marcos, P. Rabl, E. Rico, and P. Zoller, \emph{{Superconducting Circuits for
  Quantum Simulation of Dynamical Gauge Fields}}, Phys. Rev. Lett. 111 (2013),
  p. 110504,
  \urlprefix\url{http://link.aps.org/doi/10.1103/PhysRevLett.111.110504}.

\bibitem{marcos2014two}
D. Marcos, P. Widmer, E. Rico, M. Hafezi, P. Rabl, U.J. Wiese, and P. Zoller,
  \emph{Two-dimensional lattice gauge theories with superconducting quantum
  circuits}, Annals of physics 351 (2014), pp. 634--654.

\end{thebibliography}
\end{document}